\documentclass[prc,twocolumn,floatfix]{revtex4}
\usepackage{amsmath,amssymb,graphicx,bm,physics}
\usepackage{color}
\usepackage{ulem,cancel}

\newcommand{\vsrg}{V_{\text{srg}}}
\newcommand{\beq}{\begin{equation}}
\newcommand{\eeq}{\end{equation}}
\newcommand{\bea}{\begin{eqnarray}}
\newcommand{\eea}{\end{eqnarray}}
\newcommand{\ba}{\begin{align}}
\newcommand{\ea}{\end{align}}
\newcommand{\bfig}{\begin{figure}}
\newcommand{\efig}{\end{figure}}

\newcommand{\D}{\displaystyle}

\newcommand{\Wll}{W_{ll^\prime}}
\newcommand{\Vllp}{V_{ll^\prime}}
\newcommand{\phillp}{\phi_{ll^\prime}}
\newcommand{\Dllp}{D_{ll^\prime}}
\newcommand{\fmi}{\text{fm}^{-1}}  
\newcommand{\kf}{k_{\rm F}}
\newcommand{\Ef}{E_{\rm F}}
\newcommand{\vlowk}{V_{{\rm low}\,k}}

\newcommand{\MeV}{\, \text{MeV}}

\newcommand{\vekk}{\mathbf{k}}

\begin{document}

\title{Triplet Pairing in pure neutron matter} 
\author{Sarath Srinivas}
\email{saraths@physics.iitm.ac.in} 
\author{S. Ramanan}
\email{suna@physics.iitm.ac.in}
\affiliation{Indian Institute of Technology Madras, Chennai, India.} \date{\today}

\begin{abstract}
We study the zero temperature BCS gaps for the triplet channel in pure neutron matter using Similarity Renormalization Group (SRG) evolved interactions. We use the dependence of the results on the SRG resolution scale, as a tool to analyze medium and many-body corrections. In particular, we study the effects of including the three-body interactions at leading order, which appear at N2LO in the chiral EFT, as well as that of the first-order self-energy corrections on the zero temperature gap. In addition we also extract the transition temperature as a function of densities and verify the BCS scaling of the zero temperature gaps to the transition temperature.  We observe that the self-energy effects are very crucial in order to reduce the SRG resolution scale dependence of the results, while the three-body effects at the leading order do not change the two-body resolution scale dependence. On the other hand, the results depend strongly on the three-body cut-off,  emphasizing the importance of the missing higher-order three-body effects. We also observe that self-energy effects reduce the overall gap as well as shift the gap closure to lower densities. 
\end{abstract}

\maketitle

\section{Introduction}
\label{sect:intro}

  

Neutron stars are dense stellar objects that are formed after gravitational collapse 
of a massive star.  The star is mainly composed of asymmetric matter, which is neutron rich with a small percentage 
of protons whose positive charge is compensated by an equal number of electrons. The 
density of matter in the star increases radially inwards reaching values greater than the nuclear saturation density at the center. The 
 models explaining the cooling rate of neutron stars 
 suggest the existence of superfluid phases of neutrons in the inner crust and outer 
 core where they form Cooper pairs
 ~\cite{ChamelHaensel2008,Gnedin:2000me,Yakovlev:2004iq,Monrozeau:2007xu,Yakovlev:2007vs,Ofengeim:2015daa,Page:2004fy,Page:2009fu,Page:2010aw}.
Pairing occurs between neutrons in the $^1S_0$ channel which is attractive for $\kf 
 < 1.7 \,\fmi$, where $\kf$ is the Fermi momentum. At 
 higher densities, pairing between neutrons occurs in the triplet channel $^3P_2-
 ^3F_2$. The density dependence of neutron pairing in the 
 different partial wave channels is not well understood and is a problem of interest 
 in low energy nuclear physics.  Pairing between 
 protons is still an open problem, as one has to deal with the low densities of protons interacting 
 in a medium of high density neutrons.

Understanding pairing in the triplet channel is very important to explain the cooling 
of neutron stars.  For example, the recent sudden cooling of Cassiopeia-A can be explained by
the existence of the superfluid neutrons in the triplet 
channel~\cite{Page:2010aw,Ho:2009mm,Heinke:2010cr}. 
By analyzing the archival data of the Chandra X-ray observatory, it was documented that the surface 
temperature of Cassiopeia-A had suddenly decreased by a factor of $\sim 3.6\%$. Although 
observational uncertainties can change this rate to  $\sim 2.9\%$~\cite{Elshamouty:2013nfa}, 
one still requires the neutrons in the triplet channel to be in the superfluid phase with a moderately high 
transition 
temperature, to make theoretical predictions match observations. These predicted rates of decline in 
the surface temperature control the width of the density profile of the transition temperature, $T_c$, as 
well the peak value of $T_c$ as a function of density. A large value of $T_c$ allows for the star to enter 
the superfluid phase early in its thermal evolution, while a moderate value allows for a rapid cooling in the 
intermediate years.  These observations motivate the recent renewed interest in the superfluid triplet 
channel of pure neutron matter~\cite{Dong:2013sqa,Maurizio:2014qsa,Ding:2015tda,Ding:2016oxp}.

The triplet channel is particularly challenging: the reason being that the gaps 
(within the BCS approximation), remain open at much higher densities, usually in 
the range, $1 \,\fmi < \kf < 3.5 \,\fmi$. We define the BCS approximation as one where the two-body
free space interaction is used as input to the BCS gap equation, while a free-spectrum is used for the 
intermediate single particle states. In the two-body sector, a typical phenomenological 
nucleon-nucleon interaction is constrained by the two-body scattering data up to $350 \MeV$ lab 
energies or equivalently up to $2.0\, \fmi$ in momentum scale. In addition to the phenomenological 
interactions, there are EFT based chiral interactions where the low-energy constants are fit to the
two-body data and these interactions have a built-in EFT cut-off that specifies the range of momenta 
for which the interactions are valid. The biggest challenge when using these standard two-body 
interactions to study pairing
in the triplet channel is that for the range of momenta corresponding to the densities where the triplet 
gaps exist, these high-precision two-body
interactions no longer yield equivalent phase shifts in free 
space~(Fig.1 in~\cite{Baldo:1998ca} and Fig.~\ref{fig-srg-ps} in this work). 
This results in model dependent gaps in the triplet channel
~\cite{Baldo:1998ca,Dean:2002zx,Dong:2013sqa,Maurizio:2014qsa,Ding:2015tda,Ding:2016oxp}. 
 The uncertainties in the predicted gap values are influenced by the fact that the input interactions 
are no longer phase-shift equivalent in free space, in addition to the missing many-body/medium 
corrections. 

While pairing in the $^1S_0$ channel has been extensively studied, for example~\cite{Hebeler2007,Dean:2002zx,HebelerSchwenk2010}, pairing in higher partial waves, especially the 
triplet channel is not very well understood. Recently, there has been renewed 
interests in this channel~\cite{Baldo:1998ca,Dong:2013sqa,Maurizio:2014qsa,Finelli:2014xqa,Ding:2015tda,Ding:2016oxp}. 
In~\cite{Dong:2013sqa}, the authors study the triplet pairing in pure neutron matter 
within the BCS framework, taking into account the effect of short-range correlation 
via the $Z$-factor, while~\cite{Ding:2015tda,Ding:2016oxp} takes into consideration both the 
short-range and long-range correlations using self-consistent Green's function techniques and Fermi Liquid 
theory respectively. In both these approaches there is a remarkable decrease in the overall magnitude of 
the angle-averaged triplet gaps as well as the gap closure shifts to lower densities. Recently, Maurizio 
et al.~\cite{Maurizio:2014qsa,Finelli:2014xqa} have studied the gap equation in the singlet and in the triplet channel for 
both 
symmetric nuclear matter and pure neutron matter using chiral interactions at N3LO and its 
renormalization group evolved low-momentum interactions (SRG) for the symmetric matter (singlet and 
triplet), while for the triplet channel in pure neutron matter, they have documented the gaps for different 
two-body interactions, including the chiral interactions and have explored the effects of adding chiral 
three-body interactions at N2LO as well as that of including an effective mass. They solve the BCS gap 
equation following the numerically stable procedure of Khodel et al.~\cite{Khodel:2000qw}. 
 
 It is customary to work in a partial wave basis and in this basis the gap depends on $(j,l,m)$, where 
 $j$ is the total angular momentum, $l$ is the orbital angular momentum and $m$ is the projection of $j
 $.  In the triplet channel the study of gaps with $m$ dependence and partial wave mixing are by 
 themselves very interesting in the context of cooling and transport.  While the transition temperature is 
 unaffected by the 
spin of the paired state, the form of the gap below $T_{c}$ very much depends on whether the pairing 
occurs in the spin singlet
or in the spin triplet channels. Such problems have been studied in the past 
in~\cite{hoffberg,tamagaki,richardson} and 
recently in~\cite{bedaque}, where the possibility of gaps with a node that lead to special collective 
modes have been discussed. 

In this study, we re-visit the problem of pairing in pure neutron matter, a close approximation to the highly 
asymmetric neutron stars, as a function of density. Our work lays special emphasis on the pairing in the triplet channel $^3P_2-^3F_2$, thereby extending the work of 
Maurizio et al.~\cite{Maurizio:2014qsa}. We use the angle-averaged approximation to the triplet gap,  along with the assumption that the different partial waves contribute in non-overlapping intervals, as a 
starting point. Although working in this approximation is a standard first step, and is a useful one numerically, we wish to emphasize that the effects of the $m$ dependence as well as allowing the different $j$ to mix could have very interesting effects~\cite{tamagaki,Khodel:2000qw}. But this is beyond the scope of the current work. 

We use SRG-evolved interactions as input to the 
triplet channel gap equation.  
SRG evolution~\cite{Furnstahl:2013oba,Bogner:2009bt} (and references within) has the advantage of 
decoupling low and high momenta via 
unitary transformations of the Hamiltonian, while preserving observables, which are the two-body phase 
shifts and the deuteron binding energy in the two-body sector as a function of a flow parameter $s$, 
where $s^{-1/4}$ has the dimensions of momentum.  
In the canonical implementation of SRG, the evolution drives the interaction matrix elements towards the 
diagonal, leaving a low-momentum block and a high momentum tail~\cite{Bogner:2006pc}. Usually one 
uses a parameter $\lambda = s^{-1/4}$, which is a measure of the spread of the off-diagonal strength 
and sets the SRG resolution scale. This identification allows one to connect to the older class of
renormalization group based low-momentum interaction, $\vlowk$, specified by the momentum 
cut-off $\Lambda$. For more details on the SRG evolution, we refer the reader to the 
appendix~\ref{sect:appen1}.

Therefore, when we compare the gaps for different 
SRG resolution scale, $\lambda$, we have by construction removed the 
contribution of the two-body phase shift inequivalence to the $\lambda$ dependence. This allows one to 
use the dependence of the gap on $\lambda$ as 
a tool to estimate the scale of the missing physics such as many-body and/or medium dependencies.  
We note that the gaps obtained for different densities and SRG resolution scales will depend on the 
bare interactions, but we wish to only study global trends across different bare interactions. In particular, 
we compare in this study the SRG resolution scale dependence trends for the N3LO EM 
500~\cite{entem_machleidt} class of SRG interactions and the AV$_{18}$~\cite{AV18} class of SRG 
interactions so as to have examples from an EFT inspired interaction as well as a phenomenological one.
We add that the N3LO EM 500 interaction has a chiral cut-off of $500\, \MeV \sim 2.5 \, \fmi$. As a 
result, one needs to be careful while interpreting the results for the chiral interaction as $\kf$ approaches 
values close to the cut-off. In this work, we make a more conservative estimate for the build-up of the 
errors and interpret results using the N3LO interactions beyond $\kf \sim 2.0 \, \fmi$ with caution.

We augment the N3LO EM 500 bare as well as its SRG evolved counterparts by the leading 
order chiral three-body force, that occurs at N2LO in the chiral expansion. The leading order unevolved 
three-body force is regulated by an exponential regulator, parameterized by a three-body cut-off, 
$\Lambda_{\text{3NF}}$.  The three-body interaction is added to the two-body chiral interaction as a 
density dependent two-body force, obtained by integrating the third particle over the filled Fermi 
sea~\cite{HebelerSchwenk2010}. By using a leading order unevolved three-body force, we have
left out higher-order three-body effects as well as that of the induced three-body. 
Our motivation for the three-body study undertaken in this work is to primarily test the
approximation of adding unevolved leading order three-body forces as a density dependent two-body 
force for the range of densities relevant to the triplet gaps in pure neutron matter. In addition, the current 
prescription 
for adding the leading order three-body force, gives an additional three-body cut-off. Therefore one can
use the dependence on the three-body cut-off as a tool to estimate the higher-order three-body 
corrections, in addition to studying the effect of the three-body force to the triplet pairing gaps. 
There have been several studies in the past for the triplet channel of pure neutron matter where the 
two-body forces have been augmented by a density dependent phenomenological three-body 
force~\cite{Zhou:2004fz,Zuo:2008zza}. It has been observed in these studies that the addition of the 
three-body force enhances the triplet pairing gap, which is what we see in our work as well. However, we 
would like to emphasize that our study here with the leading order chiral three-body force in only 
exploratory and by no means complete. 

We solve the angle-averaged gap equation using the numerically stable procedure outlined by Khodel et 
al.~\cite{Khodel:2000qw,Maurizio:2014qsa}. Alternatively, one can extract the angle-averaged gap by picking out the poles 
of the in-medium $T$ matrix through the Weinberg eigenvalue method~\cite{Weinberg} developed in~\cite{Ramanan:2007bb} for the singlet channel. As a proof of principle, we generalize the eigenvalue method for the 
triplet channel and show that the gaps obtained are identical to the angle-averaged ones got by solving 
the gap equation in the triplet channel. The Weinberg eigenvalue method can be extended to finite 
temperatures as in~\cite{Ramanan:2013mua}, where it was applied to obtain the transition 
temperature for the $^1S_0$ channel. In this work, we extract the transition temperature as a function of 
$\kf$ for the triplet channel. 

This paper has been organized as follows. In section~\ref{sect:BCS-gap} we discuss how the gap 
equation in the partial wave basis can be solved in a numerically stable way, recapitulating the approach 
presented in~\cite{Khodel:2000qw,Maurizio:2014qsa} and we present the method of the Weinberg 
eigenvalues in order to extract the zero temperature angle-averaged gaps. In 
section~\ref{sect:Tc} we generalize the eigenvalue method to finite temperatures.
We motivate the advantage of using SRG-evolved low-momentum interactions and present the main 
results of the paper and detail the higher-order corrections that we wish to include in section~\ref{sect:results}. We summarize our results and list out our long-term goals in section~\ref{sect:summary}. Throughout this paper we work in units where $c =1$ and $\hbar^2/m_N = 1$. 
\section{The BCS pairing gap}
\label{sect:BCS-gap}
It is well known that an attractive interaction between fermions 
favors the formation of Cooper pairs, which then condense leading to an instability of the normal ground 
state.  Since the $NN$ interaction has attractive and repulsive pieces, one expects pairing in different partial wave channels as a function of density. The BCS gaps at zero temperature can be extracted by numerically solving the BCS gap equation 
in the respective partial wave channels~\cite{Khodel:2000qw,Maurizio:2014qsa}. 

One can alternatively pick out the poles of the $T$ matrix equation for energies around the Fermi 
surface~\cite{Ramanan:2007bb}. This is done by looking at the eigenvalue equation for $V G_0(E)$, 
where $G_0(E)$ is the two-particle non-interacting Green's function.  The poles of the $T$ matrix are signaled by the eigenvalues of $V G_0(E)$ approaching $1$. Therefore if one interprets the formation of Cooper pairs as the 
formation of bound states at the Fermi surface with complex energies, then the imaginary part yields the 
gap~\cite{Ramanan:2007bb}.
 
In the following subsections we present both the approaches for completeness. 

\subsection{The Gap equation and its numerical solution}
 \label{khodel}

The BCS gap equation that allows the estimation of the energy gap is given by:
\begin{equation}
\Delta(\vekk) = -\sum_{\vekk^\prime} \bra{\vekk} V \ket{\vekk^\prime} \frac{\Delta(\vekk^\prime)}
{2E(\vekk^\prime)},
\label{eq:gap}
\end{equation}
where $E(\vekk)^2 = \xi(\vekk)^2 + \Delta(\vekk)^2$ and $\xi(\vekk)$ is the single particle 
energy measured from $\Ef$, the Fermi energy, i.e., $\xi(\vekk) = e(\vekk) - \Ef$. For simplicity, we assume a free spectrum for the single particle energies to begin with and will consider corrections to this assumption in the later sections of this work. We are interested in the value of the gap at the Fermi surface as a function of $\kf$.

Working in a partial wave basis,  the gaps in the triplet channel, depend on $j$, $l$ and $m$~\cite{Khodel:2000qw}. 
Upon angle averaging, we have the following equation:
\begin{equation}
\Delta_l(k) = \sum_{l'} \frac{(-1)^{N}}{\pi}\int_{0}^{\infty} q^2dq V_{ll'}(k,q) \frac{\Delta_{l'}(q)}{E(q)},
\label{eq:gap_coup}
\end{equation}
where $N = 1 + \frac{(l - l')}{2}$ and  $E(k)^2 = \xi(k)^2 + D(k)^2$ and $\xi(k) = 
e(k) - \Ef$. In addition, following~\cite{Maurizio:2014qsa}, we have also assumed in Eq.~(\ref{eq:gap_coup}), that the gaps with different $l$ and $j$ do not overlap and hence we have  
$D(k)^2 = \sum_l\Delta_l(k)^2$ $= \Delta_1(k)^2 + \Delta_3(k)^2$ for the $^3P_2-^3F_2$ channel. Setting $l=l^\prime$ in 
Eq.~(\ref{eq:gap_coup}), we retrieve the gap equation for the uncoupled 
channels in the partial wave basis.  We note that the intermediate states have momenta in the range
$[0, \infty)$. However, in practice, the two-body interactions have non-zero matrix elements up to some maximum 
momentum,  $k_{\text{max}}$, which in turn sets the range of $q$. For notational simplicity, the limits on the intermediate state momenta will be henceforth suppressed in our discussion.
In order to solve the gap equation in a numerically 
stable way, we resort to a quasi-linear method as in~\cite{Khodel:2000qw} which was used recently by 
Maurizio et al.~\cite{Maurizio:2014qsa,Khodel:2000qw}. We begin by defining an auxiliary potential 
$\Wll(k, k^\prime)$ such that
\beq
  \Wll(k, k^\prime) = \Vllp(k, k^\prime) - {\mathit v}_{ll^\prime} \phillp(k) \phillp(k^\prime),
  \label{eq:aux}
 \eeq
 where $\phillp(k) = \D\frac{\Vllp(k, \kf)}{\Vllp(\kf, \kf)}$ and ${\mathit v}_{ll^\prime} = \Vllp(\kf, \kf)$ and 
 by construction the auxiliary potential in Eq.~(\ref{eq:aux}) vanishes if $k$ or $k^\prime$ lies on the Fermi 
 surface. The gap equation becomes:
 \begin{multline}
 \Delta_l(k) - \sum_{l^\prime} \D\frac{(-1)^N}{\pi} \int q^2 dq\, \Wll(k, q) \D\frac{\Delta_{l^\prime}(q)}{E(q)} \\
   = \sum_{l^\prime} \Dllp \phillp(k),
  \end{multline}
  and the coefficients satisfy,
  \beq
    \Dllp  = \D\frac{(-1)^N}{\pi} {\mathit v}_{ll^\prime} \int q^2 dq\, \phillp(q) \D\frac{\Delta_{l^\prime}(q)}{E(q)}.
    \label{eq:Dll}
   \eeq  
  The gap is defined as,
  \beq
    \Delta_l(k) = \sum_{l_1\, l_2} D_{l_1l_2} \chi^{l_1l_2}_l(k),
    \label{eq:gap_sep}
   \eeq
   and
   \bea
     \chi^{l_1l_2}_l(k) &-& \sum_{l^\prime} \frac{(-1)^{N}}{\pi} \int q^2 dq \, \Wll(k, q) \D\frac{\chi^{l_1l_2}_{l^\prime}(q)}
     {E(q)}\nonumber\\ 
     &=& \delta_{ll_1} \phi_{l_1l_2}(k),
     \label{eq:chi}
    \eea
    where $\delta_{ll_1}$ is the Kronecker delta.  By construction, $\chi^{l_1 l_2}_l(\kf) = \delta_{l,l_1}$ for any value of $l_{2}$,  since the potential $W_{ll^{\prime}}(\kf,q) = 0$ and $\phi_{{l_{1} l_{2}}}(\kf) = 1$. We write the energy denominator of Eq.~(\ref{eq:chi}) as 
    $E(q) = \sqrt{\xi^2(q) + \delta^2}$ where $\delta$ is a scale factor and the final result is independent of 
    the choice of $\delta$~\cite{Maurizio:2014qsa}. Eqs.~(\ref{eq:Dll}),~(\ref{eq:gap_sep}) and~(\ref{eq:chi}) are solved self-consistently to yield the gaps (or  the angle-averaged gaps for the triplet channel). 
The numerical advantage that is gained by the method of separation by Khodel et al. is that the 
singular part  of the gap equation, where the singularity arises for small values of the gap as one 
approaches the Fermi surface, is separated from the gap equation via the function $\chi^{l_1 l_2}_k(k)$. 
Further, as in (Eq.~(\ref{eq:chi})), these functions involve integrals over the auxiliary potential, which by construction go to zero on 
the Fermi surface. As a result, the functions $\chi^{l_1 l_2}_l(k)$ are insensitive to the quantity $\delta$ 
used as a first guess while solving Eqs.~(\ref{eq:Dll}),~(\ref{eq:gap_sep}) and~(\ref{eq:chi}) self-consistently.

\subsection{Stability Analysis}
\label{stability}

As an alternative to solving the gap equation, one can also look for the poles of the in-medium $T$ 
matrix. The idea is to view pairing as a non-perturbative phenomena that leads to the instability of the 
normal ground state, resulting in the divergence of the particle-particle ladder series.  The poles of the $T$ matrix are located by studying the eigenvalue equation for 
$G_0(E) V$, which is the operator that is iterated in the Born series expansion of the $T$ matrix, that is:
\begin{eqnarray}
  T(E) &=& V + VG_0(E)T(E) \nonumber \\
  &=& V + V G_0(E) V + V G_0(E) V G_0(E) V \nonumber \\ 
  &&\mbox{} + \cdots,
\end{eqnarray}
where $G_{0}(E)$ is usually the two-particle non-interacting free space Green's function. 
Therefore, if one picks a basis where $G_0(E) V$ is diagonal, i.e. 
\begin{equation}
  G_0(E) V \ket{\Gamma}= \eta(E) \ket{\Gamma},
  \label{eq:wein-eig}
\end{equation}
then the Born series expansion for the $T$ matrix becomes:
\begin{equation}
    T(E) = V(1 + \eta (E) + \eta (E)^2 + \cdots ),
  \end{equation}
which converges if $|\eta(E)| < 1$. We can immediately see that if $E$ is a true bound state of the potential, 
the eigenvalue equation, Eq.~(\ref{eq:wein-eig}), becomes the Schr\"{o}dinger's equation for the bound state 
and therefore for that energy $E$, the corresponding eigenvalue equals $1$.

The eigenvalues of the operator $G_0(E) V$, referred to as the 
Weinberg eigenvalues in the literature, allow one to track the sources of non-perturbative physics in the 
interaction $V$, which show up as eigenvalues larger than $1$, hence rendering the Born series divergent~\cite{bogner2006}. One can also use the eigenvalues as in~\cite{abrikosov,dickhoff-book,Ramanan:2007bb} to 
determine the momentum independent pairing gap.  

Pairing leads to an instability in the normal ground state resulting in the divergence of ladder diagrams, 
which should 
be reflected in the divergence of the Weinberg eigenvalues as $E \rightarrow 2 \Ef$. Close to the Fermi 
surface, one 
needs to consider hole-hole scattering in addition to particle-particle scattering. Therefore, in order to 
extract the 
pairing gap using the Weinberg eigenvalues, one replaces the two-particle free space Green's function 
by the in-medium non-interacting particle-particle hole-hole Green's function at zero center of mass
momentum. When one studies the eigenvalues of the operator $G_0(E) V$, it is seen that as a function 
of $E$, the eigenvalues diverge as 
$E \rightarrow 2 \Ef$~\cite{abrikosov,dickhoff-book,Ramanan:2007bb}. With the view that the pairing 
instability is due to the formation of bound states of Cooper pairs which then condense, we note that the 
two-body non-interacting particle-particle hole-hole Green's function can 
only accommodate the new bound state (Cooper pair) on the imaginary axis as the real axis has the 
particle-particle 
continuum for $E > 2\Ef$ and the hole-hole continuum for $E < 2\Ef$~\cite{dickhoff-book}. As a result, if 
one replaces the $E$ in the energy denominator of the Green's function by $2\Ef + i E_0$ and adjusts 
$E_0$, then value of $E_0$ for which $|\eta(2\Ef + i E_0)| = 1$ gives the pairing gap~\cite{Ramanan:2007bb}. 

For the uncoupled channels, one solves the following eigenvalue equation in the partial-wave basis:
\begin{multline}
  \D\frac{2}{\pi} \int q^{2} dq \,V_{ll}(k,q) \bigg[\D\frac{\theta(q - \kf)}{E - q^{2} + i \epsilon} \\
  - \D\frac{\theta(\kf - q)}{E - q^{2} - i \epsilon} \bigg] \Gamma_{l}(q, E) = \eta_{l}(E) \Gamma_{l}(k, E)
  \label{eq:uncoup_wein}
  \end{multline}
 where $E = 2\Ef + i E_0$.  We then dial in complex energies and pick out the value $E_0$ such that $|\eta_{l}(2\Ef + i E_0)| = 1$.

This procedure is called the stability analysis and was employed to extract the $^1S_0$ pairing gap at zero temperatures~\cite{Ramanan:2007bb}.
 
 In order to extract the gaps for the coupled channels, Eq.~(\ref{eq:uncoup_wein}) is generalized as follows:
 \begin{multline}
  \D\frac{2}{\pi} \sum_{l^\prime}\int q^{2} dq V_{ll^\prime}(k,q) \bigg[\D\frac{\theta(q - \kf)}{E - q^{2} + i \epsilon} \\
  - \D\frac{\theta(\kf - q)}{E - q^{2} - i \epsilon} \bigg] \Gamma_{l^{\prime}}(q, E) = \eta_{l}(E) \Gamma_{l}(k, E).
  \end{multline}
Once again $E = 2\Ef + i E_0$ and one dials $E_0$ searching for an eigenvalue such that $|\eta_{l}(2\Ef + i E_0)| = 1$ analogous to the uncoupled case. 

We will see in sect~\ref{sect:results} that the zero temperature gaps  extracted for the $^1S_0$ and the 
$^3P_2-^3F_2$ channels are equivalent to the corresponding gaps obtained from the gap equation. 
Therefore the stability analysis provides an alternate method to extract the pairing gaps. 

We also note that the stability method can be extended to include beyond BCS corrections as 
well as to finite temperatures (see section.~\ref{sect:Tc}). In principle, the numerical cost for the 
eigenvalue search is comparable to solving the BCS gap equation. Both are very sensitive to the 
distribution of the momentum grid points, especially around $\kf$.  Although the numerical costs are 
the same for the two methods, we find
that solving the BCS gap equation through the method of Khodel et al. is very robust numerically for
small values of the gap by construction. In the stability method, the smallest possible gap is limited by 
the denominator becoming singular around $\kf$. Hence one might observe that the gaps open and 
close abruptly compared to the gaps 
obtained by solving the BCS equation through the method of Khodel et al.  In addition, we also note that the
Khodel method is applicable for \emph{any} interaction~\cite{Khodel:2000qw}, while the eigenvalue 
method works very well for interactions that are \emph{soft} such as, SRG evolved interactions at lower 
SRG resolution scales, chiral interactions etc~\cite{Ramanan:2007bb}. For such soft interactions, the 
large eigenvalue close 
to $\Ef$ is associated with pairing, but for interactions where repulsive pieces exist, such as bare 
AV$_{18}$ in the $^1S_0$ or in the $^3P_2-^3F_2$ channels the stability method is more involved, as the 
large values for $|\eta(E)|$ could arise due to both the pairing instability as well as due to the presence of 
short-range pieces that renders an ``in-medium'' Born series divergent. 
In this case, one needs to impose additional constraints on the phase of the eigenvalue. 

In the following section, we
discuss the extension of the Weinberg eigenvalue method to finite temperatures.
\subsection{Transition temperature via Weinberg Eigenvalues}
\label{sect:Tc}

The method of obtaining the pairing gaps at zero temperatures via the Weinberg eigenvalues can be easily extended to finite temperatures as in~\cite{Ramanan:2013mua}. This is done by replacing the zero temperature Green's function by the finite temperature counterpart evaluated at zero center of mass momentum:
\begin{equation} 
G_0(\vekk, \omega) =
  \frac{1 - 2\,f(\xi(\vekk))}
    {\omega - 2 \,\xi({\vekk})  + i \eta}.
\label{eq:two_body_gf_retarded}
\end{equation}
where $f(\xi) = 1/(e^{\beta\xi}+1)$ is the Fermi-Dirac distribution
function and $\omega = E-2\mu$, $\xi(\vekk) = e(\vekk)-\mu$ are the energies measured from the chemical potential.
While working in a partial wave basis with the free-particle spectrum for the single particles, one obtains the following Weinberg eigenvalue equations at finite temperature applicable to the uncoupled and coupled channels respectively:
\begin{multline}
\frac{2}{\pi} \int q^2 \,dq\,V_{ll}(k,q) 
  G_0(q,\omega)\Gamma_{l}(q,\omega)\\ 
  = \eta_{l}(\omega) \Gamma_{l}(k,\omega)\,,
\label{eq:wein_kspace_uncoup}
\end{multline}
and
\begin{multline}
\frac{2}{\pi} \sum_{l^\prime}\int q^2 \,dq\, V_{ll^\prime}(k,q) 
G_0(q,\omega)\Gamma_{l^\prime}(q,\omega)\\ 
  = \eta_l(\omega) \Gamma_l(k,\omega)\,,
\label{eq:wein_kspace_coup}
\end{multline}
where we have suppressed the dependence on $\mu$ and $T$ in 
Eqs.~(\ref{eq:wein_kspace_uncoup}) and~(\ref{eq:wein_kspace_coup}) for notational simplicity.
At finite temperatures, the largest eigenvalue that equals $1$ is picked for $\omega = 0$ for a given $\mu$ and $T$. This temperature then is the transition temperature $T_c$ for a given $\mu$. The condition:
\begin{equation}
  |\eta_{l}(\omega = 0, \mu, T)| = 1,
  \label{eq:thouless}
\end{equation}
corresponds to the Thouless criterion for the critical temperature for a non-local interaction~\cite{Ramanan:2013mua,thouless}. While this technique has been used to get $T_c$ for the uncoupled channel in~\cite{Ramanan:2007bb}, in this work, it has been extended for the coupled channel. In the next section, we present our results for both the $^1S_0$ and the $^3P_2-^3F_2$ channels.

\section{Numerical Results at zero and finite temperatures}
\label{sect:results}

We begin by studying the pairing gap at zero temperatures for the $^1S_0$ and $^3P_2-^3F_2$ channels using both the techniques of solving the BCS gap equation and the stability analysis. While the zero temperature $^1S_0$ gaps within the BCS approximation are not new, it serves to benchmark the codes and the techniques. 

 \begin{figure}[ht]
   \includegraphics[width=3.25in, clip = true]{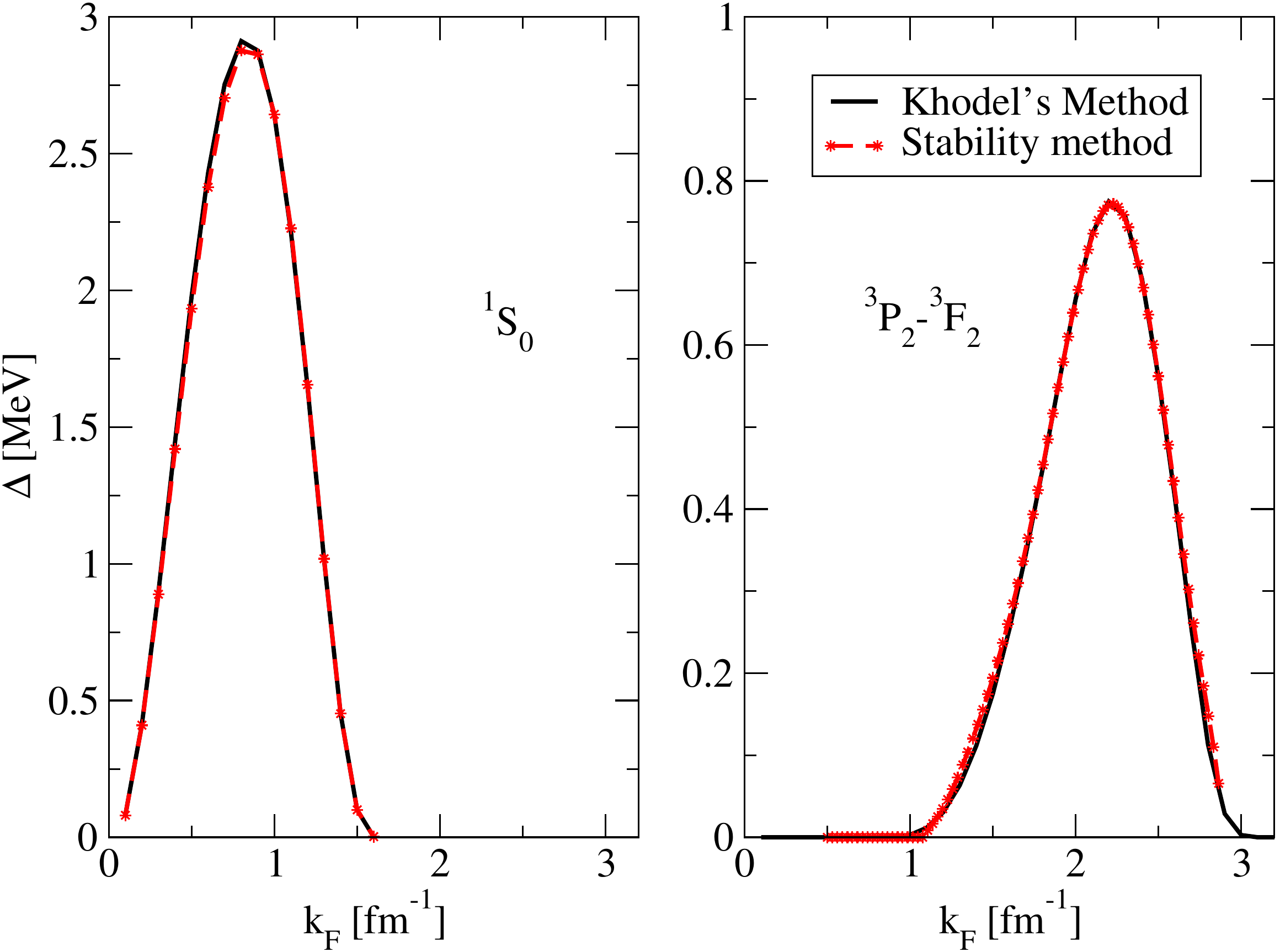}
 \caption{(Color online) Zero temperature gap for the singlet (left) and the triplet (right) obtained from 
 chiral N3LO potential~\cite{entem_machleidt}.  The $^{1}S_{0}$ gaps uses SRG evolved interaction at 
 $\lambda = 2.0 \,\fmi$, while the $^{3}P_{2}-^{3}F_{2}$ gaps has the unevolved interaction as input. 
 Obtaining the gaps for higher resolution scale for the $^{1}S_{0}$ channel using the stability method is 
 complicated  due to the presence of short range components in the interaction, that is usually softened 
 by the RG evolution~\cite{Ramanan:2007bb}.}
    \label{fig-ztgaps}
\end{figure}

Fig.~\ref{fig-ztgaps} shows the zero temperature momentum independent gaps for the $^1S_0$ and 
$^3P_2-^3F_2$ channels for pure neutron matter as a function of $\kf$. We have used the chiral N3LO 
potentials~\cite{entem_machleidt} as the input for the gap equation. The gap equation is first angle-
averaged and then solved in a numerically stable way due to Khodel (see section~\ref{khodel}) (solid 
lines in Fig.~\ref{fig-ztgaps}). We also show the zero temperature gaps obtained via the stability analysis 
(dashed-lines) and we note that the two methods agree. Our results for the triplet channel agree with 
those found in the literature for the chiral N3LO interactions~\cite{Ding:2015tda}. 

\begin{figure}[ht]
   \includegraphics[width = 3.25in, clip = true]{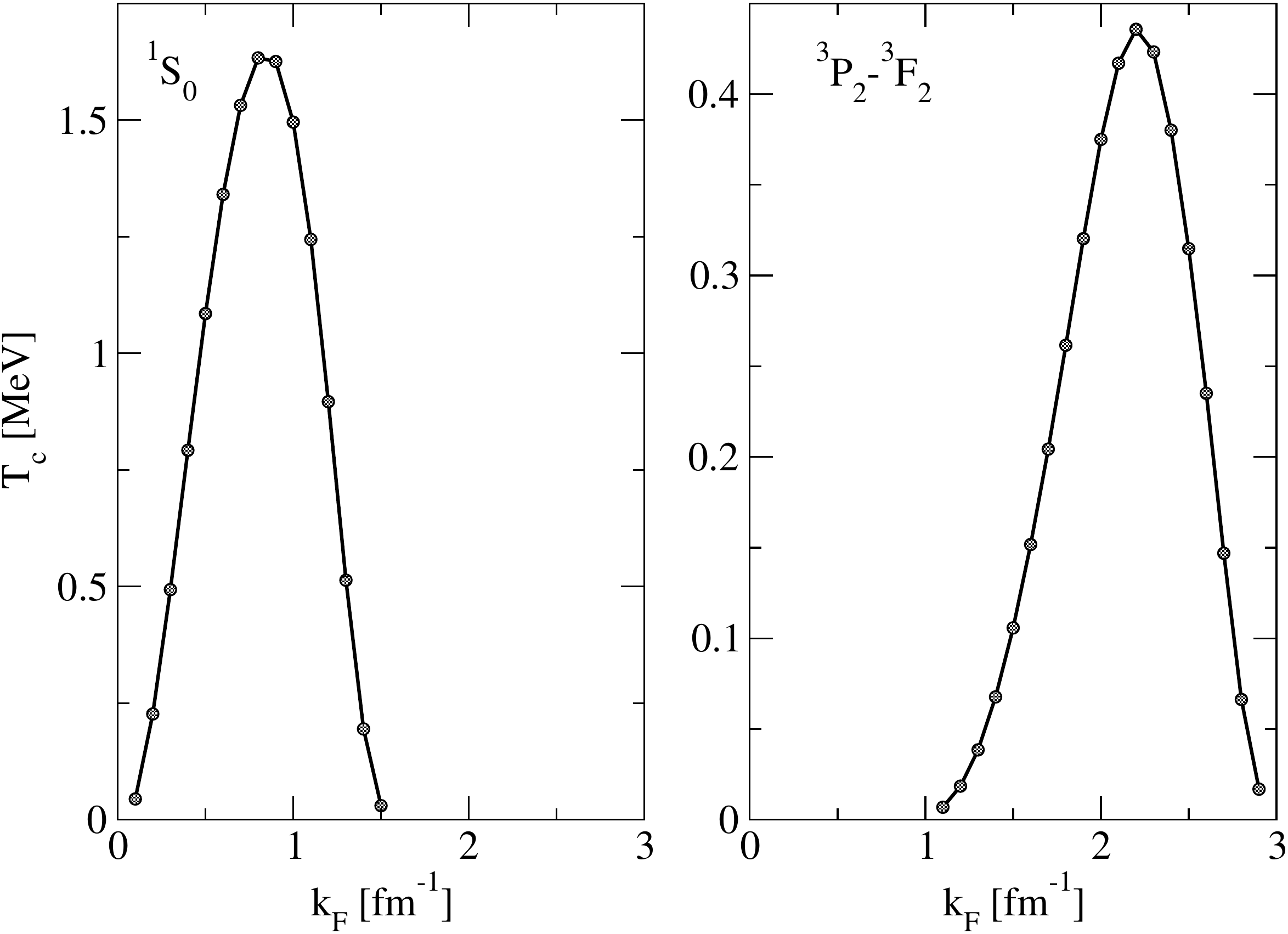}
  \caption{Transition temperature as a function of $\kf$ for the singlet (left) and the triplet 
  (right) obtained from the chiral N3LO potential~\cite{entem_machleidt}.
  Again, $T_{c}$ for the $^{1}S_{0}$ channel uses 
  the SRG evolved interaction with $\lambda = 2.0 \,\fmi$, while the triplet channel uses unevolved 
  interaction as input.}
  \label{fig-ftTc}
\end{figure}

Fig.~\ref{fig-ftTc} shows the transition temperature as a function of $\kf$ for the 
chiral N3LO interaction obtained via the Thouless criterion in Eq.~(\ref{eq:thouless}). The ratio of the zero 
temperature gap to the transition temperature follows the BCS result~\cite{fetter}, i.e.
\begin{equation}
  \D\frac{\Delta(\kf)}{T_c} \sim 1.76
\end{equation}

As outlined in the introduction, the strategy we adopt for the SRG resolution scale ($\lambda$) 
dependence study of the triplet channel gaps is to use the SRG-evolved interactions as input for the gap 
equation. We wish to re-emphasize that the main motivation for using the SRG-evolved interactions for a 
given bare interaction is that the bare phase shifts are preserved and one can then attribute the $\lambda
$ dependence to the missing medium/many-body contributions. However, we will see the dependence 
on the bare interaction if we compare the results between different interactions, for example between 
N3LO and AV$_{18}$ and their corresponding SRG-evolved interactions. An alternative to this will be to 
compare the different models without the SRG evolution and include many-body/medium corrections 
until the results are model independent. But we wish to take advantage of the systematics that the EFT 
approach offers and hence we use the SRG-evolved interactions and study the resolution scale 
dependence for a given bare interaction. 

\begin{figure}[t]
  
\begin{center}
  \includegraphics[angle = 0, width = 3.5in, clip = true]{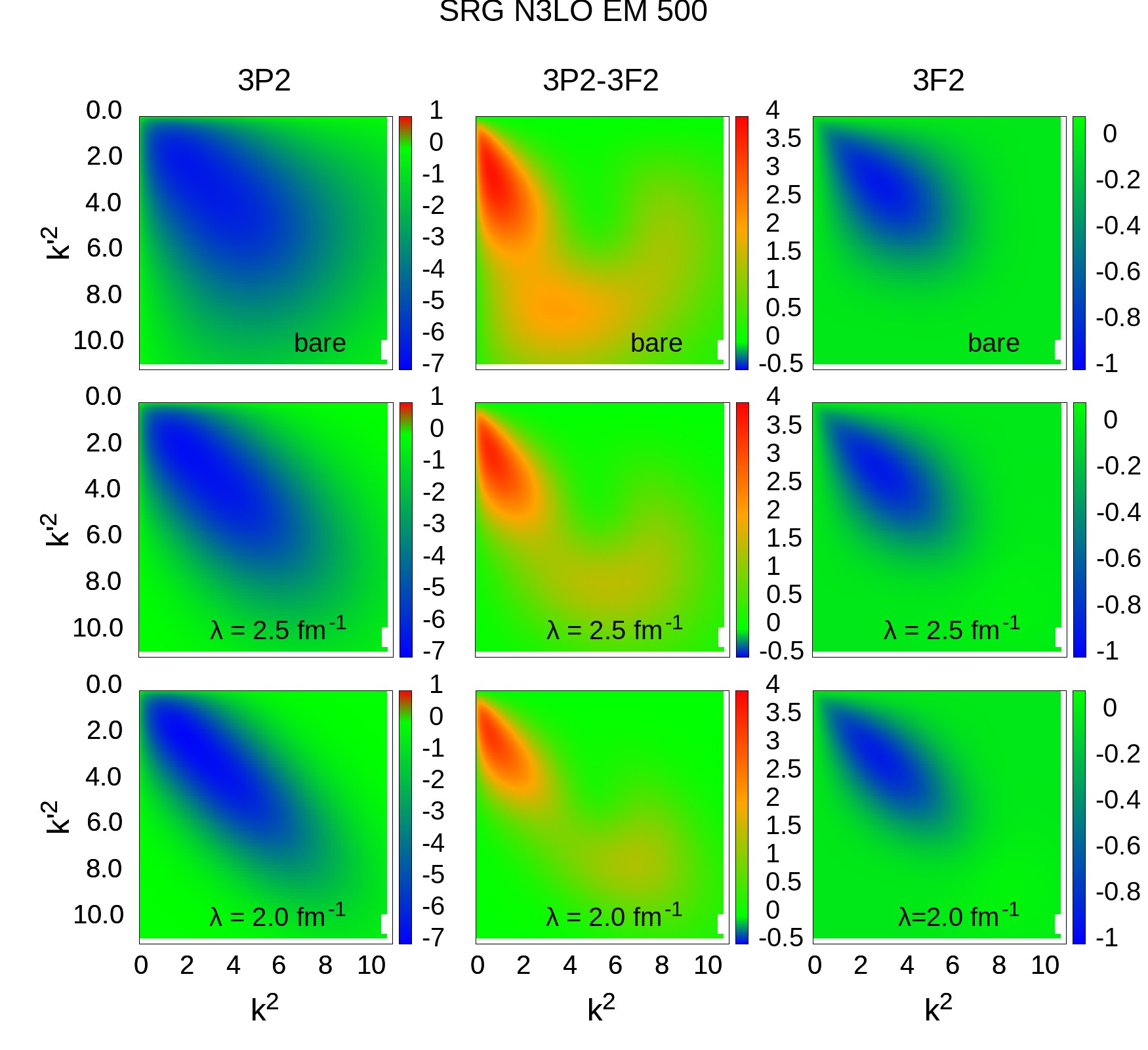}
  \includegraphics[angle = 0, width = 3.5in, clip = true]{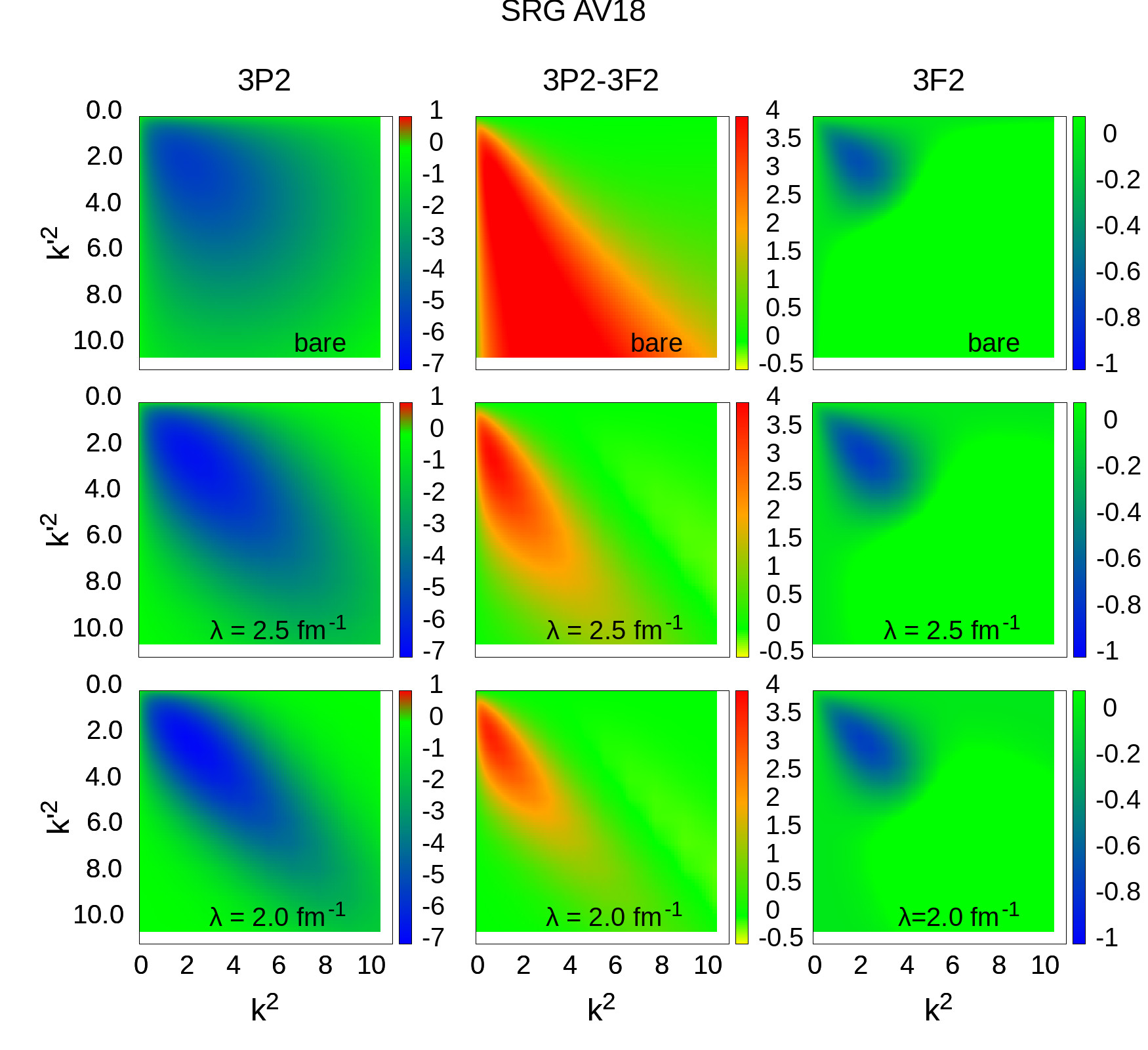}
\end{center}

  \caption{(Color online) SRG-evolved $^3P_2$$-$$^3F_2$ interactions as a function of 
  $k^2$ and $k^{\prime\,^2}$ for the N3LO EM 500~\cite{entem_machleidt} interaction (top panel) and 
  the AV$_{18}$ interaction (bottom panel). Note that the evolution drives the matrix elements towards 
  the diagonal as function of the parameter $\lambda$.}
  \label{fig-srg-evol}
\end{figure}

\begin{figure}[t]
  \begin{center}
  \includegraphics[angle = 0, width = 3in, clip = true]{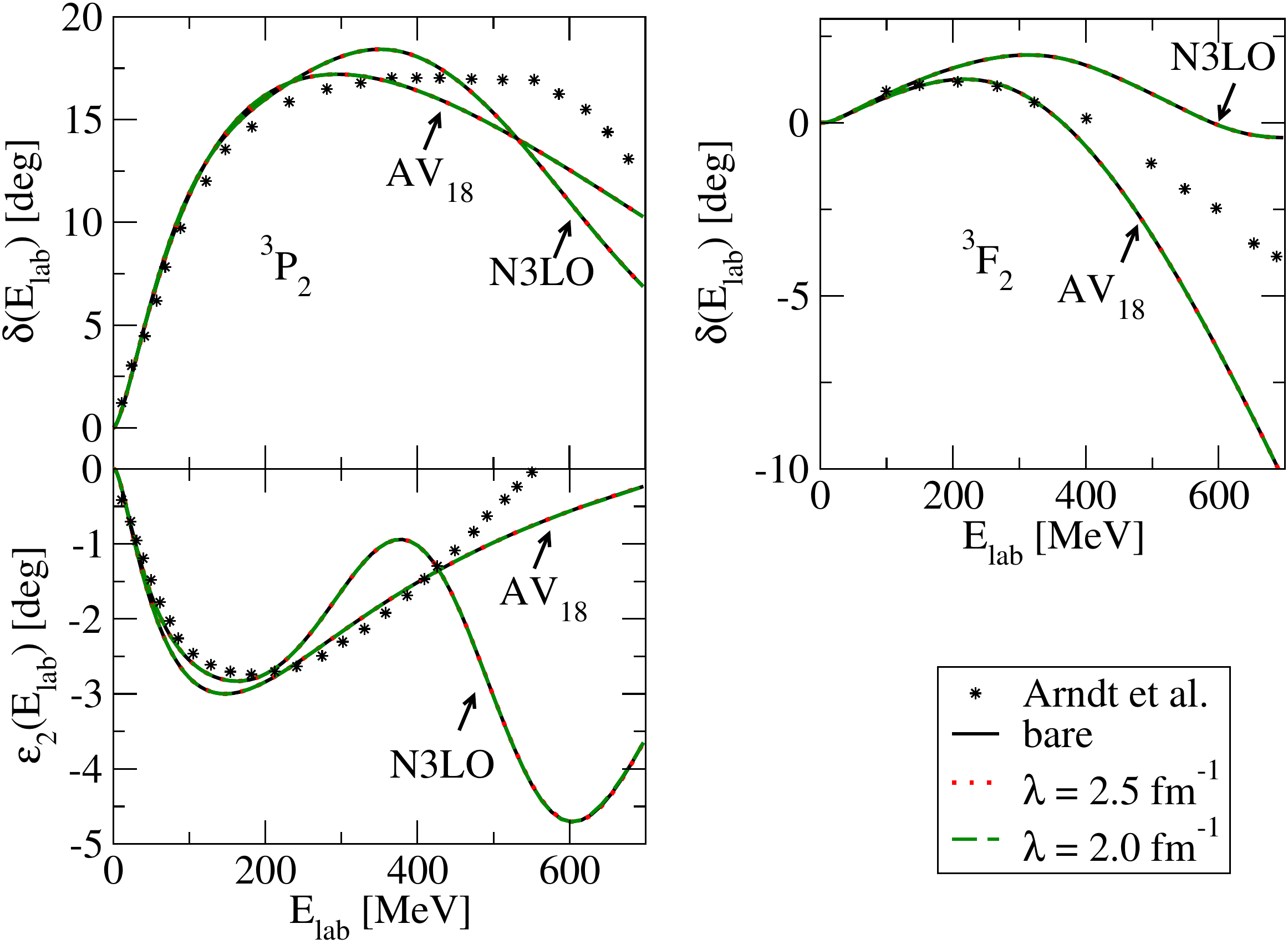}
  \end{center}
  \caption{(Color online) Comparing the phase shift in the triplet channel for the N3LO EM 500
  ~\cite{entem_machleidt} and the AV$_{18}$~\cite{AV18} interaction. Note that the SRG evolution in 
  each case preserves the phase shift as seen by the lack of dependence on the SRG resolution scale. 
  For comparison, we have also included the partial wave analysis of the experimental phase shifts by 
  Arndt et al.~\cite{Arndt:1997if}}
  \label{fig-srg-ps}
\end{figure}

\begin{figure}[h]
  \includegraphics[angle = 0, width = 3.25in, clip = true]{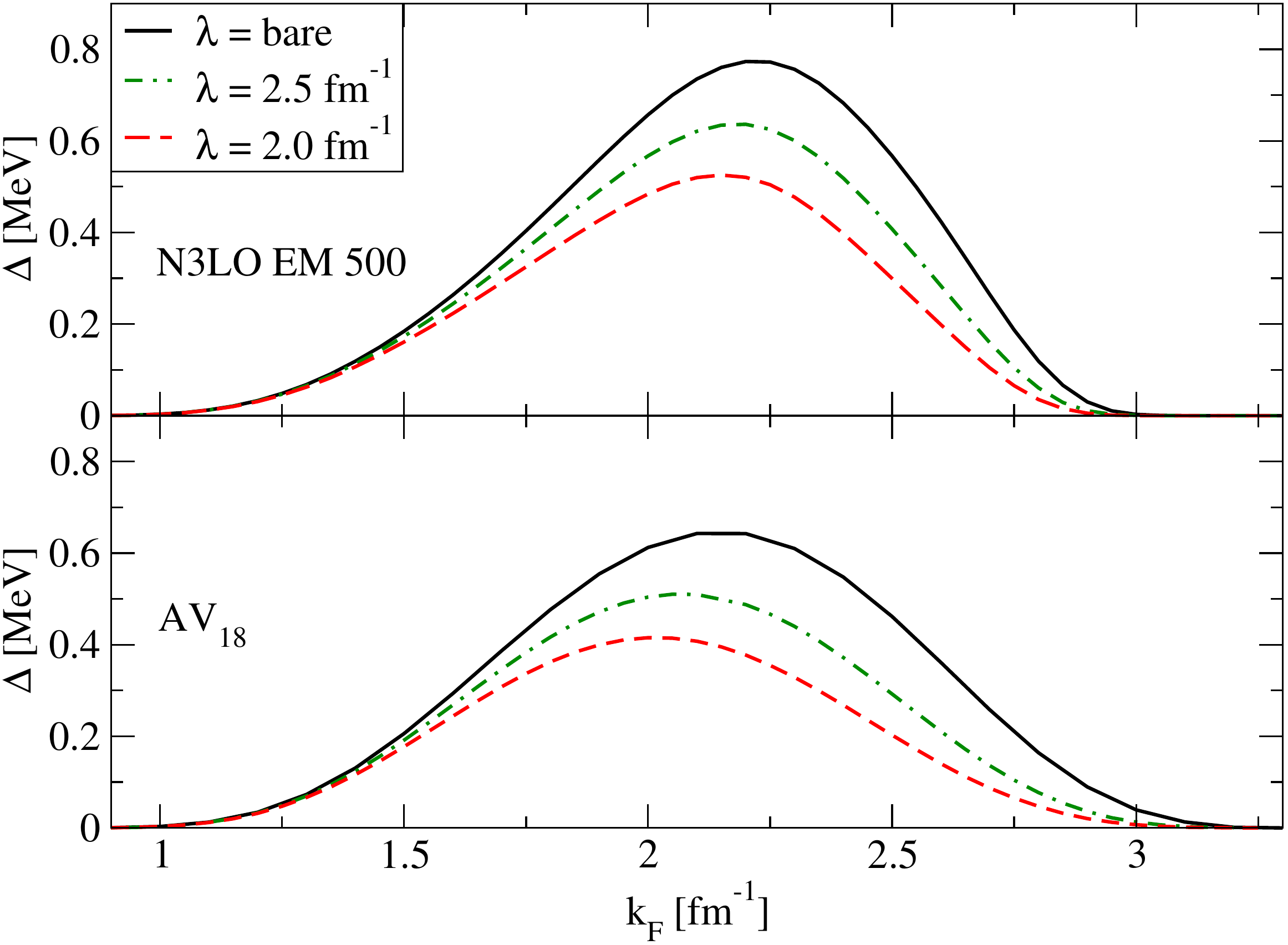}
  \caption{(Color online) Zero temperature gaps obtained from N3LO EM $500$ and AV$_{18}$ for 
  different SRG resolution scales. The top panel has the N3LO EM 500 bare and SRG-evolved interactions 
  as inputs. The bottom panel has the AV18 bare and SRG-evolved interactions for the same $\lambda$ 
  values.}
  \label{fig-zeroT-gap-SRG-cutoff}
\end{figure}

We begin by revisiting the effect of the SRG evolution on the bare interaction. Fig.~\ref{fig-srg-evol} 
shows the SRG 
evolution for the N3LO EM 500~\cite{entem_machleidt} and the AV$_{18}$~\cite{AV18} interactions in 
the triplet 
channel for pure neutron matter. We see that the evolution in both the cases, drives the interaction matrix 
elements 
towards the diagonal and this has consequences on the gaps, namely, the gaps decrease as the 
resolution scale $\lambda$ decreases (see Fig.~\ref{fig-zeroT-gap-SRG-cutoff}). In addition, the 
evolution preserves the bare phase shifts as it is unitary.  This is seen in Fig.~\ref{fig-srg-ps}, where the 
phase shifts for different SRG resolution scales are identical to the bare (unevolved) phase shifts for both 
the N3LO EM 500 and the AV$_{18}$ interactions. In Fig.~\ref{fig-srg-ps}, we also show the partial wave 
analysis of the experimental $NN$ scattering data of Arndt et al.~\cite{Arndt:1997if} for comparison, 
and it is clearly
seen that the phase shifts computed using the different $NN$ interactions depart significantly from the 
the experimental data at high energies. Hence when the triplet gaps from the different $NN$ interactions 
are compared, the uncertainties in the calculated gaps also include the phase shift inequivalence that
occur at higher energies in addition to the missing many-body/medium corrections. Therefore, 
we work with the AV$_{18}$ and the chiral N3LO interactions and 
their corresponding SRG evolved counterparts and we plan to study the $\lambda$ dependence of the 
triplet gaps for a particular bare interaction. However, we note that the N3LO interaction cannot be 
trusted beyond the chiral EFT cut-off of $2.5 \, \fmi$, and conservatively, one should interpret the N3LO 
results beyond $\kf \sim 2.0 \, \fmi$ with caution.

\begin{figure}[h]
  \includegraphics[angle = 0, width = 3.25in, clip = true]{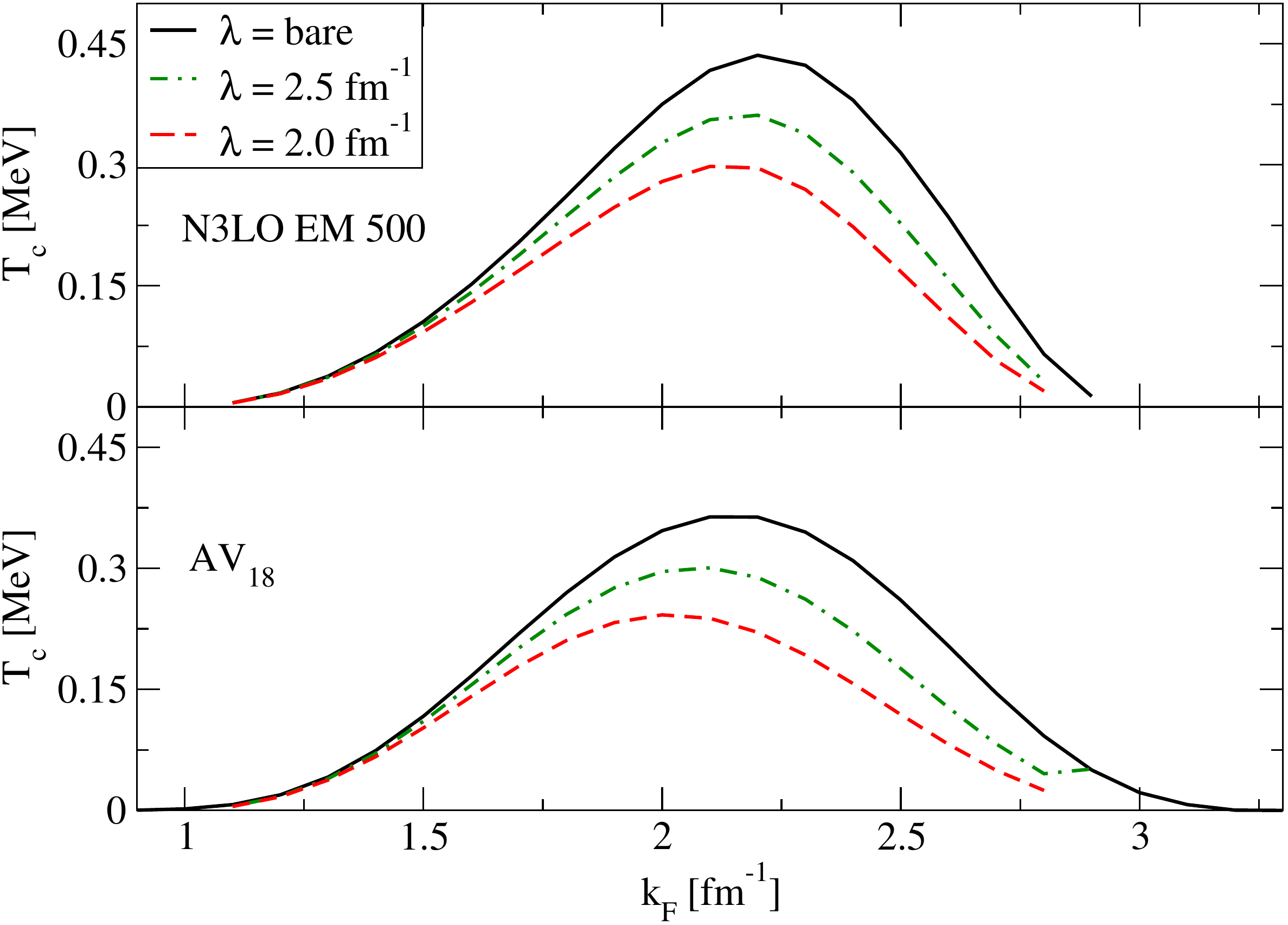}
  \caption{(Color online) Transition temperature as a function of $\kf$ for N3LO EM $500$ and AV$_{18}$ 
  for different SRG resolution scale. The top panel has the N3LO EM 500 bare and its SRG-evolved 
  interactions as inputs. The bottom panel has the bare AV18 and its SRG-evolved interactions as inputs.}
  \label{fig-ft-gap-SRG-cutoff}
\end{figure}

The triplet gaps at zero temperature for the SRG-evolved N3LO EM 500 and the evolved AV$_{18}$ 
interactions are seen in Fig.~\ref{fig-zeroT-gap-SRG-cutoff} in the top panel and bottom panel 
respectively, while Fig.~\ref{fig-ft-gap-SRG-cutoff} shows the $\lambda$ dependence of the transition 
temperature obtained via the Thouless criterion (Eq.~(\ref{eq:thouless}))$^1$. We see that for both the 
N3LO and the AV18 class of interactions, lowering $\lambda$ decreases the gap. This decrease can be 
linked to the changes the SRG evolution, determined by the generator, makes to the matrix elements, 
where the different $l\,l^\prime$ blocks are driven towards the diagonal (refer Fig.~\ref{fig-srg-evol}). 
Figs.~\ref{fig-zeroT-gap-SRG-cutoff} and~\ref{fig-ft-gap-SRG-cutoff} also document the SRG resolution 
scale dependence. We note that the results are independent of the SRG resolution scale for 
$\kf \approx 1.3 \fmi$. The same trends carry over to the transition temperature in 
Fig.~\ref{fig-ft-gap-SRG-cutoff}. The $\lambda$ dependence sets the scale of the missing 
many-body/medium corrections as a function of $\kf$. 
 
 \footnotetext[1]{For the transition temperature using the bare AV$_{18}$ interaction, there were 
 numerical  issues beyond $2.0 \, \fmi$ and the transition temperature as seen in 
 Fig.~\ref{fig-ft-gap-SRG-cutoff} for the  \emph{bare AV$_{18}$} alone is obtained from the BCS scaling. 
 However, for the SRG-evolved AV$_{18}$ interaction as well as for the N3LO (bare and evolved), the 
 Thouless criterion for the transition temperature turns out to be numerically stable.}

So far in our discussions, we use the free space interaction matrix elements (bare or equivalent SRG-
evolved interactions) for the vertex.  For the intermediate states the free particle spectrum is used. But at 
finite densities, corrections to both the vertex and the single-particle energy are important. Therefore, in 
this study we will include the $3N$ interaction as an effective density dependent $2N$ interaction as 
well as correct the free-particle spectrum with the first-order self-energy term. We  begin by reviewing 
the $3N$ corrections to the vertex. 

\begin{figure}[t]
  \includegraphics[angle = 0, width = 2in, clip = true]{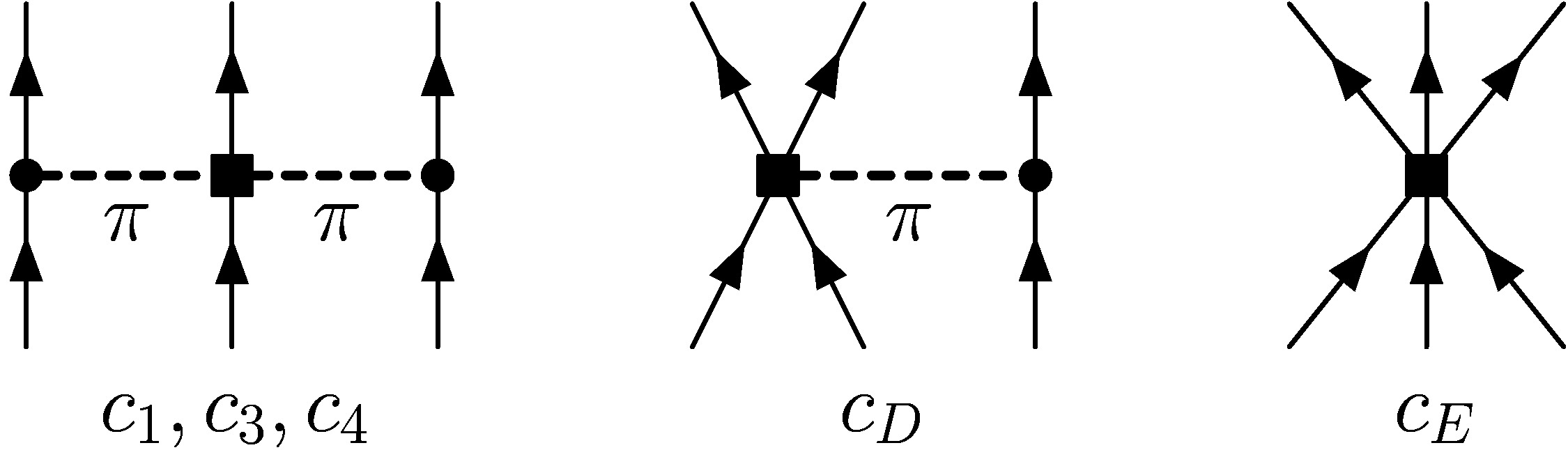}
  \caption{Leading chiral $3N$ forces at N2LO~\cite{HebelerSchwenk2010}.}
  \label{fig-chiral-3N}
\end{figure}

\begin{figure}[h]
  \begin{center}
    \includegraphics[angle = 0, width = 3.25in, clip = true]{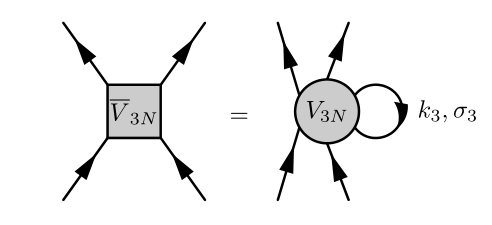}
  \end{center}
  \caption{Effective 2N generated by integrating the third particle over the states occupied in the 
Fermi sea.}
\label{fig-eff2n}
\end{figure}

The three-body force has been in the past included via an effective two-body density dependent 
interaction~\cite{Maurizio:2014qsa, HebelerSchwenk2010}. Following the previous work by Hebeler and 
Schwenk~\cite{HebelerSchwenk2010} for the $^1S_0$ channel, we include the leading order chiral 
$3N$ interactions which occur at N2LO (Fig.~\ref{fig-chiral-3N}). In pure neutron matter only the long-
range two pion exchange diagram contributes and further in this diagram only the $c_1$ and $c_3$ 
terms contribute~\cite{HebelerSchwenk2010}. We use values of $c_1 = -0.81 \,\text{GeV}^{-1}$ and 
$c_3 = -3.2 \,\text{GeV}^{-1}$~\cite{entem_machleidt,HebelerSchwenk2010}.  For the $3N$ interaction 
the following smooth regulator is used:
\beq
f_{\text{R}}(p,q) = \exp \biggl[ - \biggl(\frac{p^2+3 q^2/4}{\Lambda_{\rm 3NF}^2}\biggr)^{n_{\text{exp}}}
\biggr],
\eeq
where $p$ and $q$ are the Jacobi momenta and $\Lambda_{\rm 3NF}$ is the three-body cut-off and 
$n_{\text{exp}}$ is the parameter for the exponential regulator. We use $n_{\text{exp}} = 2$ in our study. 
Since we are including the long-range two pion force, we assume that it is not modified by the RG 
running and use the same values for $c_1$ and $c_3$ for different $\Lambda_{\rm 3N}$ cut-off.  In order 
to obtain the effective $2N$ interaction, we integrate the third particle over the states occupied in the 
Fermi sea (schematically shown in Fig.~\ref{fig-eff2n}), which is then added to the two-body interaction with appropriate symmetry factors.

For the BCS gap, the two body interaction augmented by the effective two-body term obtained from the 
$3N$ force is given by~\cite{HebelerSchwenk2010}:
\beq
  V_{\text{eff}}(k,k^\prime) = V_{\text{2N}}(k,k^\prime) + \D\frac{\overline{V}_{\text{3N}}(k,k^\prime)}{2},
  \eeq
  where $\overline{V}_{\text{3N}}(k,k^\prime)$ denotes the effective density dependent $2N$ force.
In addition to studying the dependence of the results on the SRG resolution scale, the three-body cut-off 
can be varied independently and the cut-off dependence of the results at the three-body level (which 
gives an estimate of the missing short-range three-body forces) can be analyzed.

\begin{figure}[t]
  \includegraphics[angle = 0, width = 3.25in, clip = true]{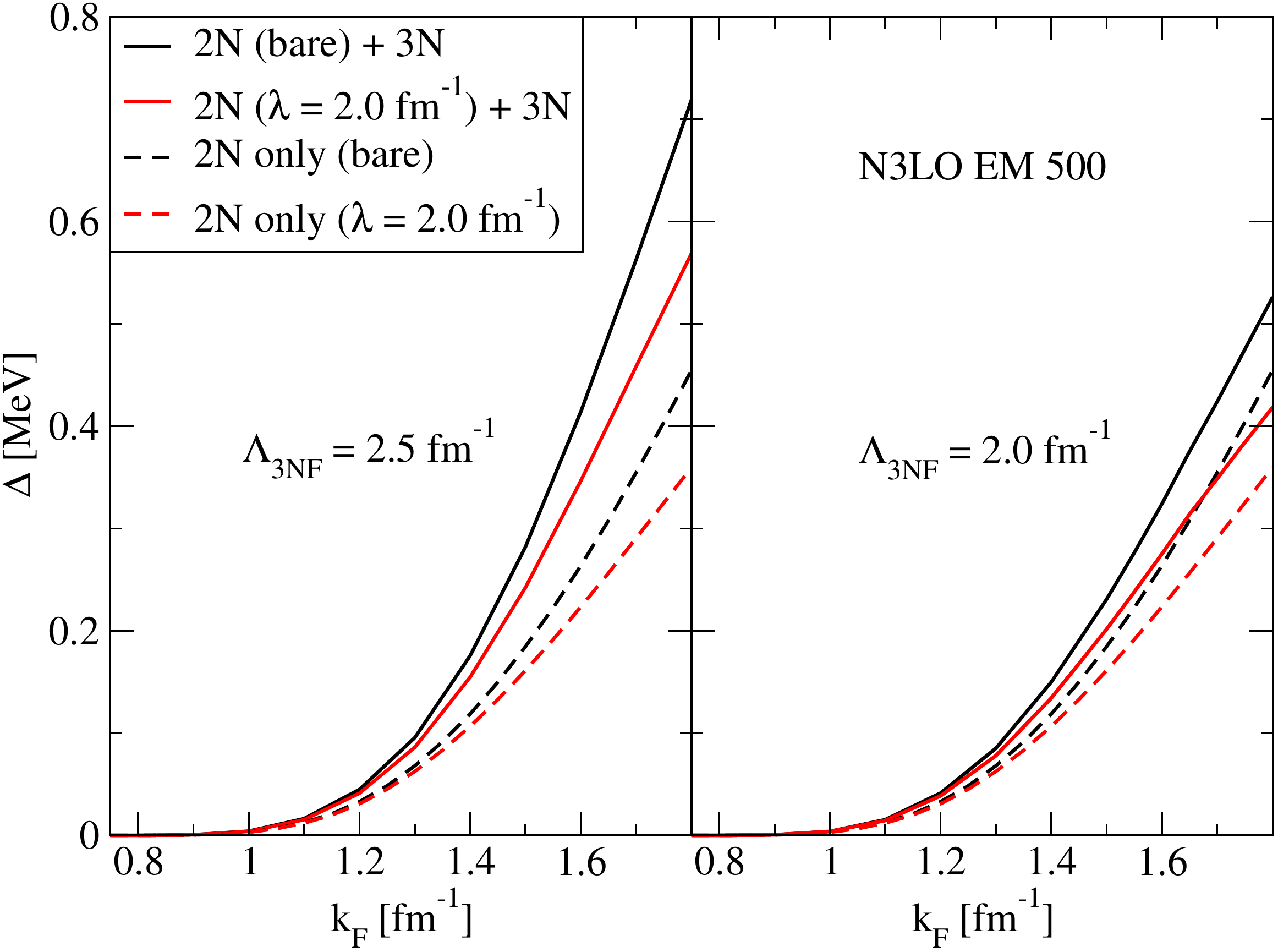}
  \caption{(Color online) Two-body SRG resolution scale dependence for the N3LO EM 500, keeping $
  \Lambda_{\text{3NF}}$ fixed when the two-body input is augmented by the effective $2N$ interaction.}
  \label{fig-2n-cutoff-dep-with-3n-fixed}
\end{figure}

Fig.~\ref{fig-2n-cutoff-dep-with-3n-fixed} shows the zero temperature gap as a function of $\kf$ when 
the input two-body interaction is augmented by the effective density dependent $2N$ force (solid 
lines). For comparison, 
the figure also shows the 2N only results (broken lines) as well. We note that for a given two-body 
resolution scale, $\lambda$, and three-body cut-off, $\Lambda_{\text{3NF}}$, the addition of the 
$3N$ force as a density dependent effective $2N$ interaction increases the triplet gaps (compare black 
solid and black broken lines). This 
increase in the gap is due to the attractive spin-orbit force that the effective $2N$ force adds to the input 
interaction~\cite{HebelerSchwenk2010}. Fig.~\ref{fig-2n-cutoff-dep-with-3n-fixed} also shows the 
spread in the two-body resolution scale for a fixed three-body cut-off. It is 
observed that independent of the three-body cut-off, the two-body $\lambda$ dependence is 
unaffected by the addition of the leading order three-body interaction. Therefore one is still 
missing important many-body corrections. In Fig.~\ref{fig-2n-cutoff-dep-with-3n-fixed}, we have 
restricted the density range to $\kf < 2.0 \, \fmi$. This limit for the density range
is due to the fact that a leading order approximation for the $3N$ interaction has been used. Hence
one cannot expect it to be valid at high densities. As mentioned in the introduction, the chiral interactions 
have a cut-off which in the case of N3LO EM 500 is around $500 \MeV \, \sim 2.5 \, \fmi$ and the 
errors build up as this value is approached. Hence we use a conservative range for $\kf$ values and  
restrict the range to $< 2.0 \, \fmi$  when the $3N$ interactions are included. We would like to add in 
addition that we have neglected the induced $3N$ forces as we use unevolved $3N$ interactions and 
these are certainly important~\cite{Hebeler:2013ri} and will be considered in a future publication.

\begin{figure}[t]
  \includegraphics[angle = 0, width = 3.25in, clip = true]{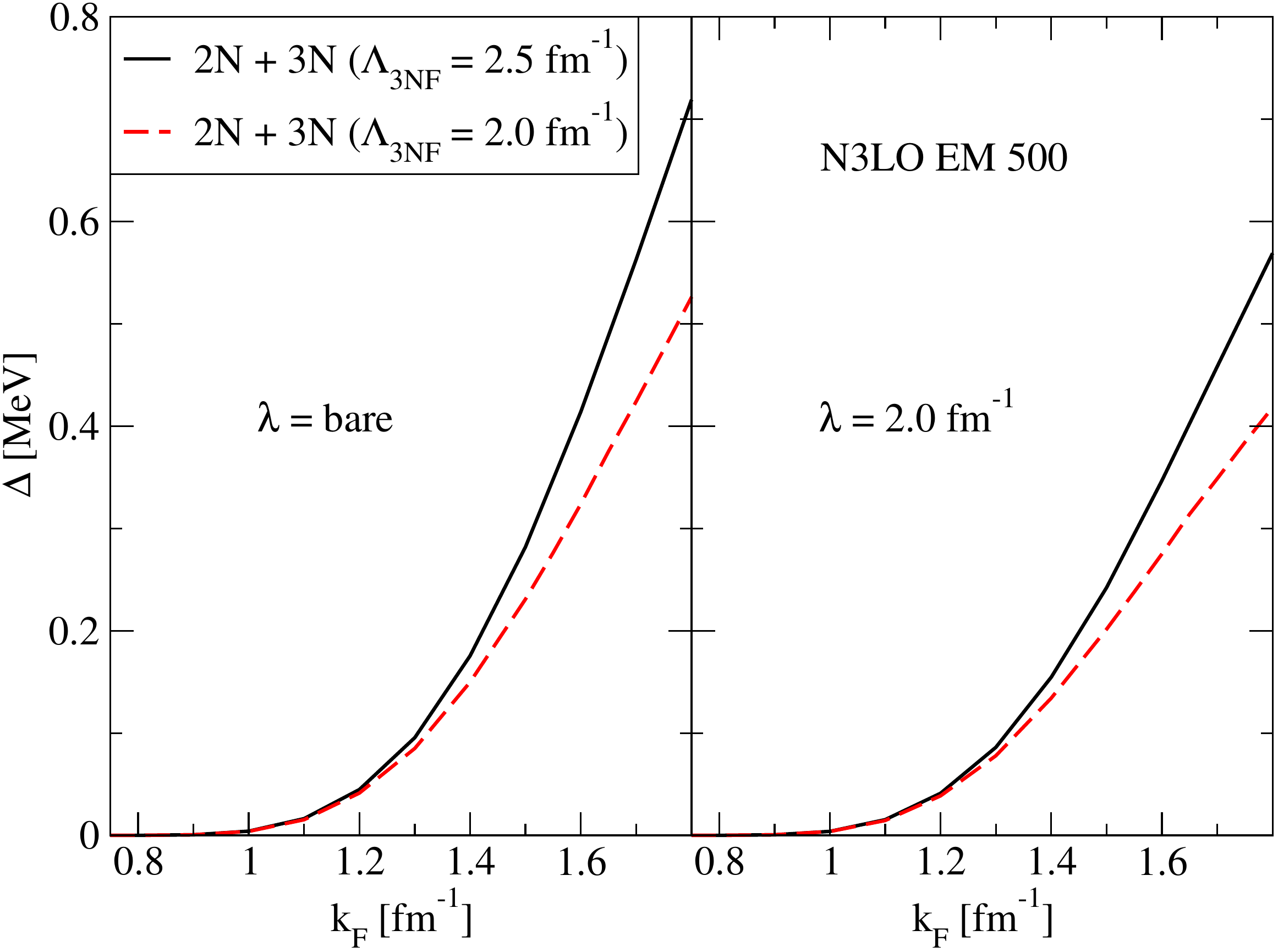}
  \caption{(Color online) Three-body cut-off dependence for the N3LO EM 500, keeping the two-body 
  SRG resolution scale $\lambda$ fixed, when the two-body input is augmented by the effective $2N$ 
  interaction.}
  \label{fig-3n-cutoff-dep-with-2n-fixed}
\end{figure}

To check the approximation made for the leading order effective three-body corrections, we compare in 
Fig.~\ref{fig-3n-cutoff-dep-with-2n-fixed} the $3N$ cut-off dependence when the $2N$ resolution 
scale, $\lambda$, is held fixed. The results show strong dependence on $\Lambda_{\rm 3N}$ thereby 
indicating that the short- and intermediate-range $3N$ effects are important for the range of densities 
considered. In fact, comparing Figs.~\ref{fig-2n-cutoff-dep-with-3n-fixed} 
and~\ref{fig-3n-cutoff-dep-with-2n-fixed}, one sees that the results are more sensitive to the $3N$ 
cut-off compared to the $2N$ resolution scale. This in turn goes back to our motivation of restricting
the densities in Figs.~\ref{fig-2n-cutoff-dep-with-3n-fixed} and~\ref{fig-3n-cutoff-dep-with-2n-fixed}.
Since the uncertainties with the $3N$ interaction is quite large within the approximation used here, we 
will not include the $3N$ corrections for the rest of the paper and will work with the $NN$-only 
vertex.

\begin{figure}[h]
  \begin{center}
    \includegraphics[angle = 0, width = 3.in, clip = true]{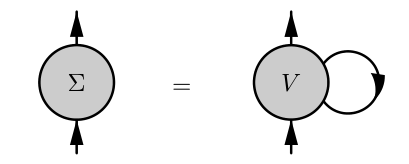}
  \end{center}
  \caption{First-order self-energy.}
\label{fig-sigma_fo}
\end{figure}

In order to correct for the single-particle spectrum we include the self-energy effects to first-order. 
Therefore, the energy $e(k)$ of the intermediate states become:
\beq
  e(k) = \D\frac{k^2}{2} + \Sigma^{(1)}(k)
  \label{eq:self-energy}
 \eeq
 where $\Sigma^{(1)}(k)$ is the static first-order contribution that is diagrammatically shown in Fig.~\ref{fig-sigma_fo}. 

The first-order term can be written in the partial wave basis as follows~\cite{HebelerSchwenk2010}:
\begin{eqnarray}
  \Sigma^{(1)}(k_1) &=& \D\frac{1}{2\pi} \int k_2^2 \,dk_2 \int d(\cos\,\theta_{\vekk_1, \,\vekk_2}) \, n_{\vekk_2} \nonumber \\ 
  &&\hspace*{-1in}\mbox{} \sum_{l,S,J} (2J+1) \ip{k_{12}/2}{V_{SllJ}|k_{12}/2} (1 - (-1)^{l+S+1}),
  \label{eq:self-energy_eqn}
\end{eqnarray}
where $n_{\vekk_2} = \theta(\kf - k_2)$ is the Fermi-Dirac distribution at zero temperature and $k_{12} = |\vekk_1 - \vekk_2|$ and we align the $z$ axis in the direction of $\mathbf{k}_1$. Since we are interested in the gaps at $\kf$, it is useful to study the effective mass $m^*$ defined as (using units $\hbar^2/m_N = 1$)
\begin{equation}
  \D\frac{m^*}{m} = \Bigg(\frac{1}{k}\, \frac{de(k)}{dk}\Bigg)^{-1}\Bigg|_{k = \kf}.
\end{equation}
The effective mass is directly related to the density of states at $\kf$ and a lower $m^*/m$ indicates 
depletion of states at $\kf$. 

\begin{figure}[t]
  \begin{center}
    \includegraphics[angle = 0, width = 3.25in, clip = true]{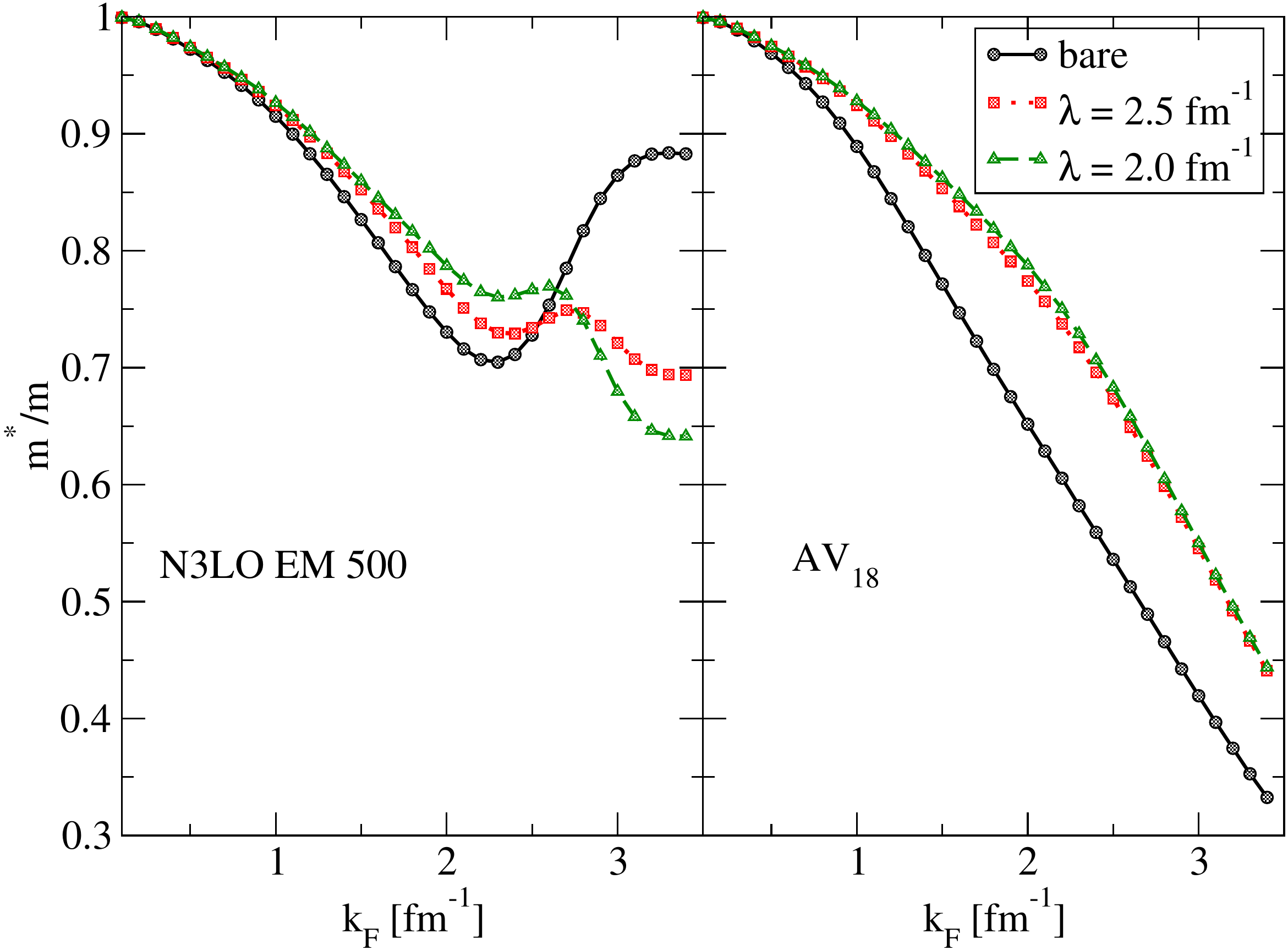}
  \end{center}
  \caption{(Color online) Effective mass using the first order self-energy term for N3LO EM 500 and AV$_{18}$ interactions for bare and different SRG resolution scales.}
  \label{fig-mstar}
\end{figure}

\begin{figure}[h]
  \begin{center}
    \includegraphics[angle = 0, width = 3.25in, clip = true]{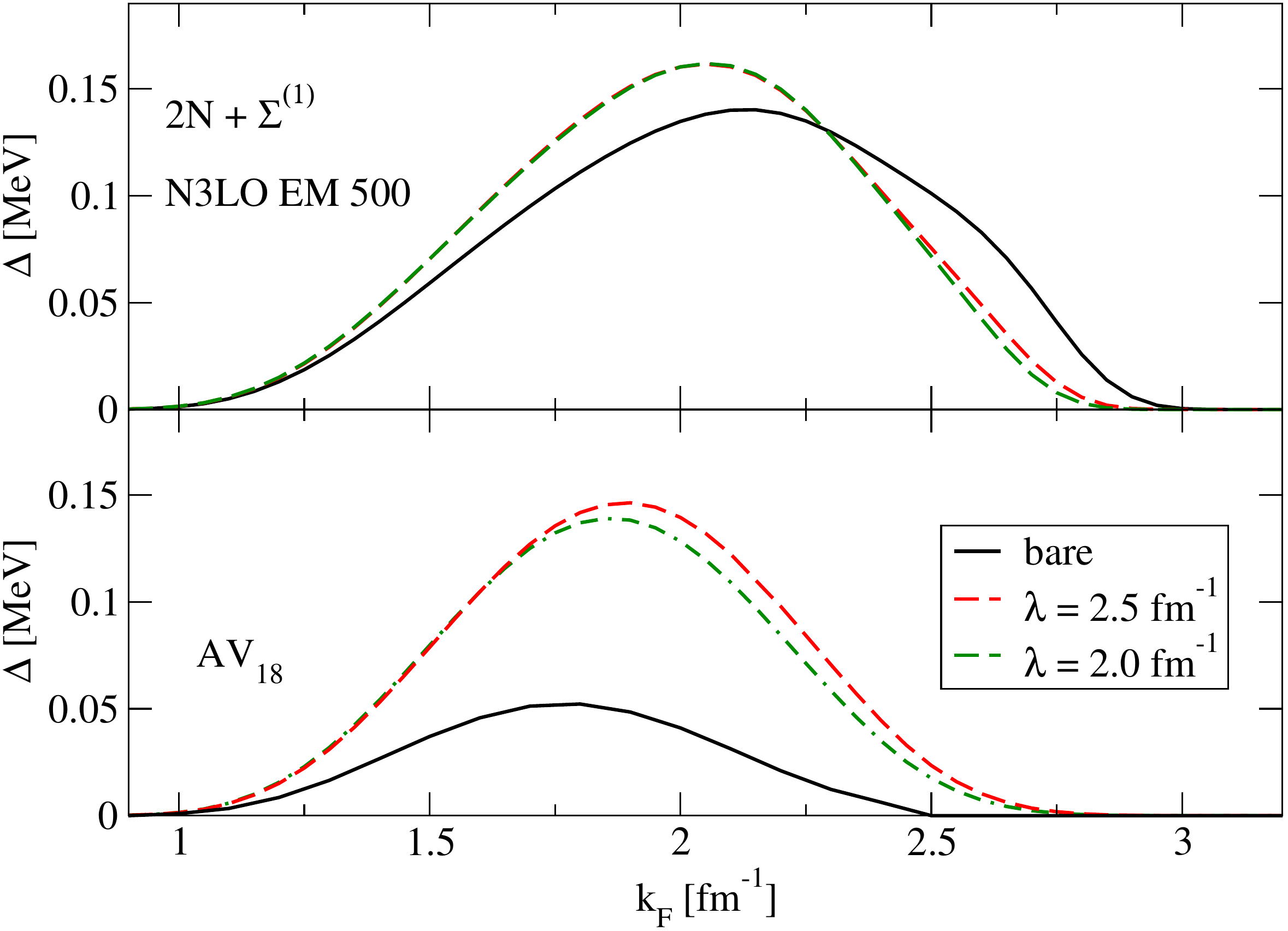}
  \end{center}
  \caption{(Color online) First-order self-energy effects on the zero temperature gap using N3LO EM 500 and AV18 respectively and the corresponding SRG-evolved interactions as inputs.}
  \label{fig-self-ener-SRG-gap}
\end{figure}

Fig.~\ref{fig-mstar} shows the $\kf$ dependence of the effective mass, where the ratio $m^*/m$ is 
plotted as a function of $\kf$ for the N3LO EM 500 (left panel) and the AV$_{18}$ (right panel) as well as 
the respective SRG evolved interactions for $\lambda = 2.5 \, \fmi$ and $2.0 \, \fmi$. The ratio of 
$m^*/m$ as a function of density decreases for both the AV$_{18}$ and the N3LO EM 500 interactions 
as a 
function of $\kf$. However,  for the N3LO interactions beyond $2.0 \, \fmi$, this ratio increases, although 
it is still less than $1$. But 
beyond  $2.0 \, \fmi$, the chiral interaction becomes unreliable as momenta/densities approach the 
chiral cut-off of $2.5\, \fmi$. This is roughly where the 
systematics with the SRG evolution scale breaks down for the N3LO as seen in Fig.~\ref{fig-mstar}. For 
the AV$_{18}$ interaction, it is not surprising that the first order approximation to the self-energy is rather 
poor as can be seen by the small values of $m^*/m$ (compare with Fig.~5 of~\cite{Baldo:1998ca}). This 
is especially true with the bare which is known be a hard interaction. For the SRG evolved AV$_{18}$ 
interactions the first order approximation for the self-energy is expected to break down at higher densities. 

\begin{figure}[h]
  \begin{center}
    \includegraphics[angle = 0, width = 3.25in, clip = true]{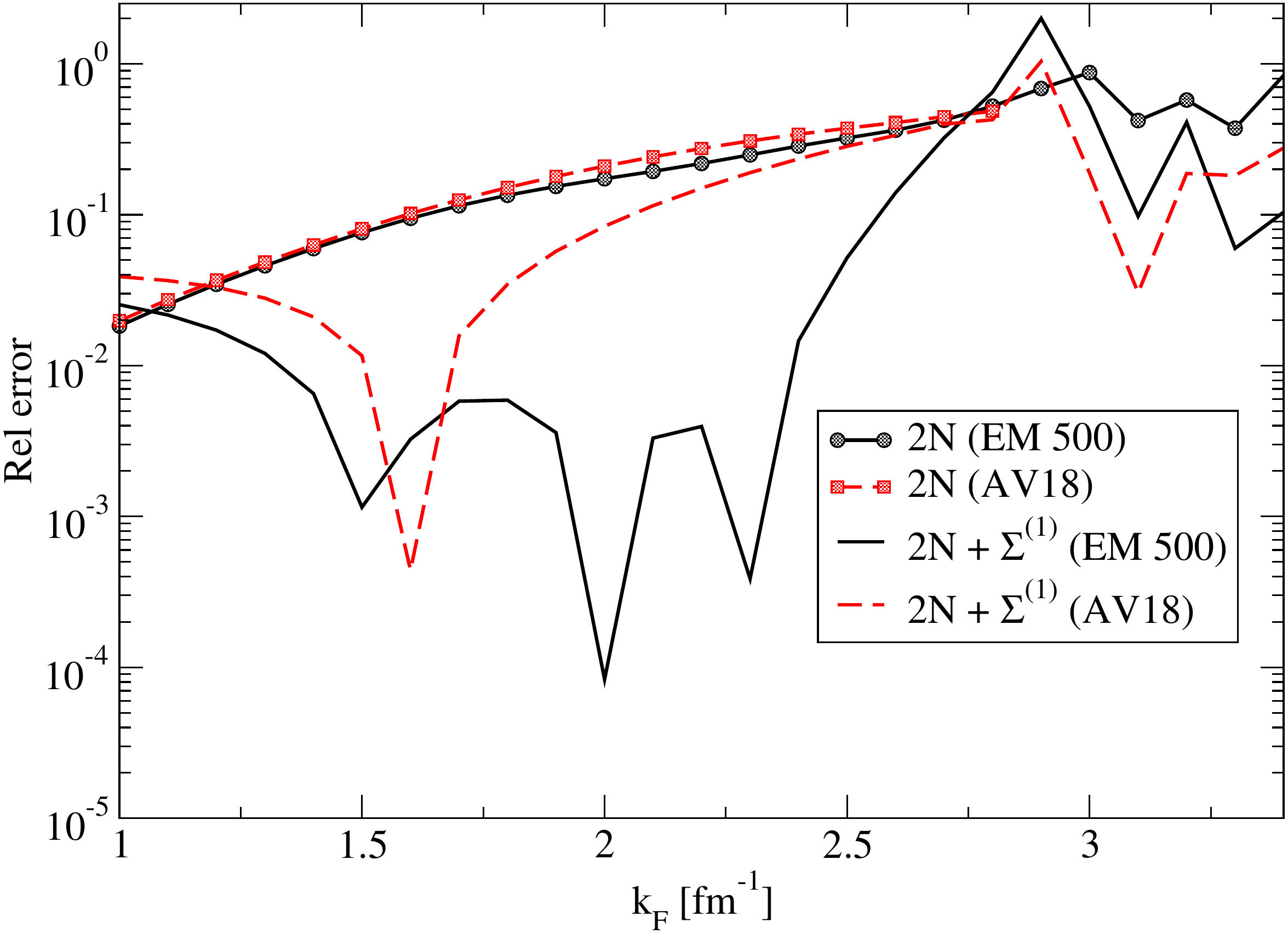}
  \end{center}
  \caption{(Color online) Relative error between the cut-offs ($\Lambda = 2.0\, \fmi$ and $\Lambda = 2.5\, \fmi$) for N3LO EM 500 and AV18 SRG-evolved interactions.}
  \label{fig-self-ener-SRG-errors}
\end{figure}

From the behavior of $m^*/m$ as a function of $\kf$, one expects that with the first order self-energy 
correction as given in Eqs.~(\ref{eq:self-energy})  and~(\ref{eq:self-energy_eqn}), the triplet gaps should 
decrease in magnitude, which is what is observed in Fig.~\ref{fig-self-ener-SRG-gap}, compared to the 
ones with the free spectrum (see Fig.~\ref{fig-zeroT-gap-SRG-cutoff}).  Further,  for the SRG evolved 
interactions, in each case, we note that the gaps increase in magnitude compared to the corresponding 
bare interaction (solid black line) in Fig.~\ref{fig-self-ener-SRG-gap}, which could be linked to the 
specifics of the SRG evolution. However, one needs to interpret the first-order self-energy correction to the bare 
interaction with caution. In particular for the bare AV$_{18}$, one sees that the gaps are very small 
compared to the SRG 
evolved counterparts. This drastic depletion of the gaps could indicate that the first order approximation 
for the self-energy for the bare AV$_{18}$ is rather poor. Similarly for the SRG evolved AV$_{18}$ 
interactions, one observes that the $\lambda$ dependence sets in around $\kf \sim 1.7 \, \fmi$, thereby 
indicating the need for higher-order corrections for the self-energy.  On the other hand, with the N3LO 
EM 500, while the 
differences between the bare interactions and the SRG evolved ones are not as striking as the 
AV$_{18}$, one expects the first order correction to the self-energy to breakdown at higher $\kf$s, 
especially for the bare. The SRG evolved N3LO interactions are soft and one observes that the 
results have minimal cut-off dependence for $\kf$ in the range $[1.0\, \fmi - 2.2\, \fmi]$, although one
should interpret the N3LO results beyond $\kf$ of $2.0\, \fmi$ with caution. 

In order to better track the cut-off dependence, we study the relative errors in the zero temperature gaps 
as a function of $\kf$ for the $2N$-only results with (lines without symbols) and without (lines with 
symbols) the first order self-energy correction in Fig.~\ref{fig-self-ener-SRG-errors} for the SRG evolved
interactions. The relative error is
obtained by taking the normalized differences of the zero temperature gaps at two different SRG 
resolution scales for the N3LO and AV$_{18}$ interactions. Not taking into account the densities where 
the gap opens (or any accidental cancellations in the relative error), we note 
that there is an overall reduction in the relative error for densities less than $2.0 \,\fmi$ when a first order 
self-energy correction is used (lines without symbols) compared to the case where a free spectrum is 
used for the intermediate states. At higher 
densities, higher order corrections to the self-energy become important. The reduction in the relative 
errors  for the AV$_{18}$ interaction is not on the same scale when  
compared with the chiral interaction and this could be due to the differences between the 
phenomenological and EFT based interactions. Therefore, from Figs.~\ref{fig-self-ener-SRG-gap} 
and~\ref{fig-self-ener-SRG-errors}, one can conclude that the addition of self-energy effects, although 
at first order, decreases the resolution scale dependence and is an important correction.

\begin{figure}[t]
  \begin{center}
    \vspace*{0.2in}
    \includegraphics[angle = 0, width = 3.25in]{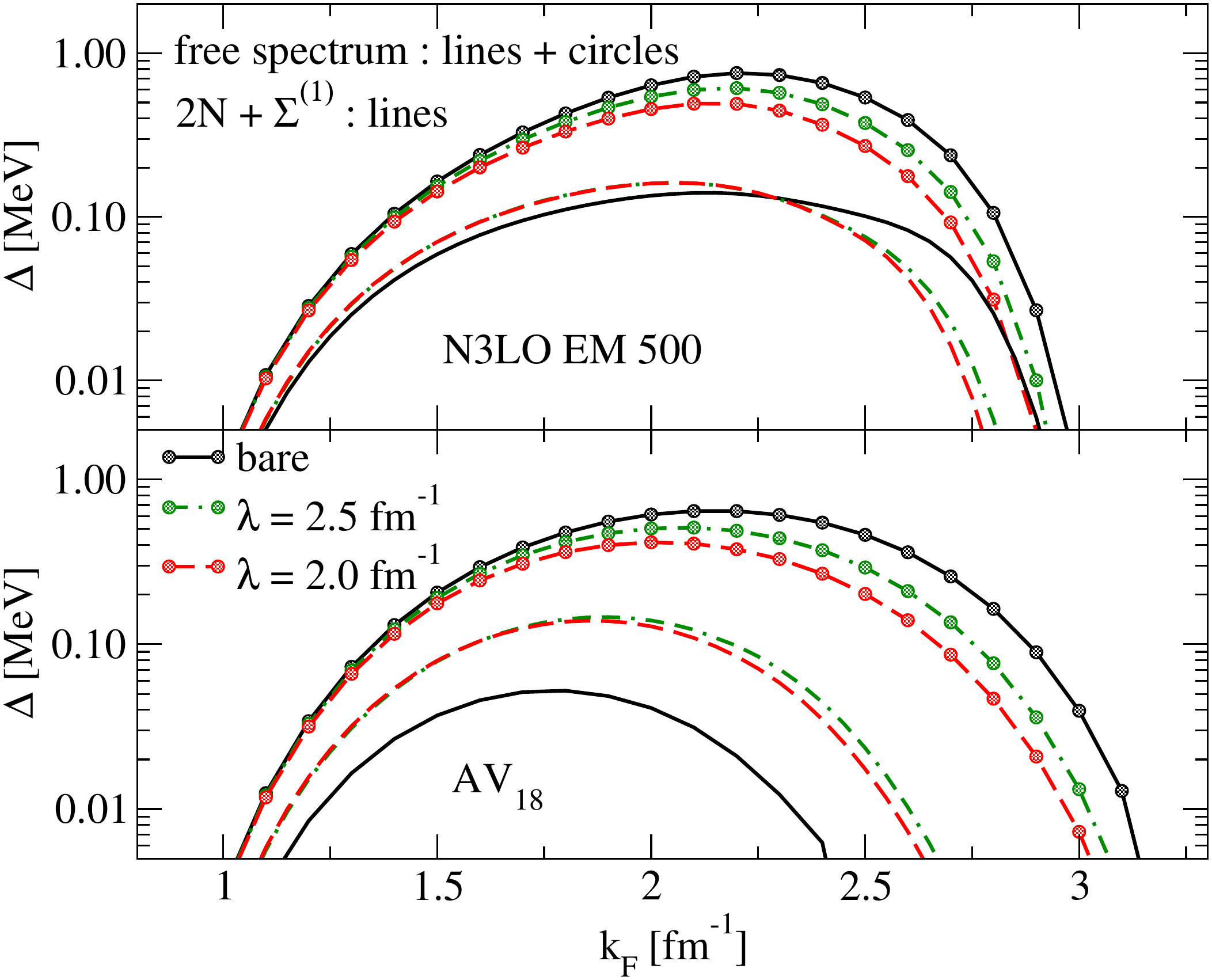}
  \end{center}
  \caption{(Color online) Effect of first-order self-energy versus the free spectrum for N3LO (top panel) and AV$_{18}$ (bottom panel). The log-scale details the effects of SRG running on the gaps as well as the closure.}
  \label{fig-self-ener-srg-av18-compare}
\end{figure}

Another striking feature of including the self-energy correction is that the gap closure shifts to lower 
densities as seen in Figs.~\ref{fig-self-ener-SRG-gap} and~\ref{fig-self-ener-srg-av18-compare}. The 
shift is similar to the one observed when short range correlations (SRCs) are included in the interaction 
(Fig.~10~\cite{Ding:2016oxp}). 
The effects of self-energy versus the free-spectrum is examined in detail in 
Fig.~\ref{fig-self-ener-srg-av18-compare}, where we use a log-scale to better document the effects of 
the cut-off as well to understand the impact of the first order self-energy correction on the gap closure. 
Ding et al.~\cite{Ding:2015tda,Ding:2016oxp} observe a lowering of the gap and a shift in the gap 
closure to lower densities when they include short-range correlations within a self-consistent Greens 
function theory. For the N3LO EM 500 one cannot conclude much about the closure, but for the 
AV$_{18}$ interaction, we note that the gap closes at lower $\kf$ when the self-energy at first-order is 
included, although the closure is at a higher density compared to the Ding et al results.  We note and 
emphasize that a first order correction to the self-energy may not be sufficient to comment without 
ambiguities about gap closures.  It would be interesting to study the effects of second order self-energy 
as well as higher order effects systematically on the gaps in the triplet channel.

\begin{figure}[h]
  \begin{center}
    \includegraphics[angle = 0, width = 3.25in]{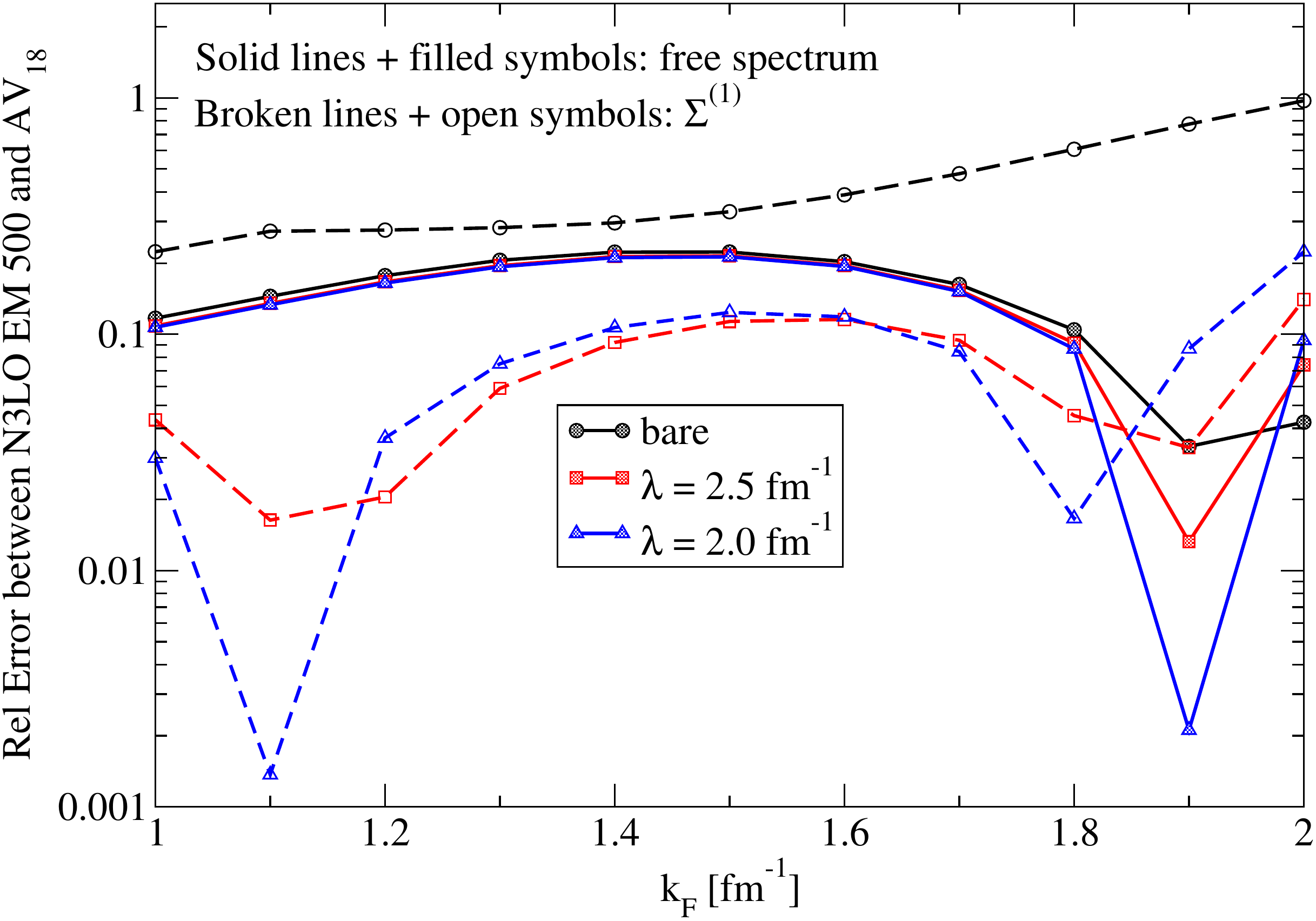}
  \end{center}
  \caption{(Color online) Relative error between N3LO EM 500 and AV$_{18}$ interactions for different cut-offs including the bare interaction with and without the self-energy corrections. The solid lines with filled symbols are the relative errors for the free spectrum, while the broken lines with the open symbols include self-energy corrections to first-order.}
  \label{fig-model-dep}
\end{figure}

We have been emphasizing that the triplet gaps obtained thus far will depend on the interaction used in 
the BCS gap equation as the densities are very high and the interactions are not phase-shift equivalent in 
free space. 
In addition, one cannot expect the free-space interaction to 
completely describe the pairing in the triplet channel. As an attempt to track the dependence of the 
results 
on the free-space interaction, we study the relative error, which are normalized differences between gaps 
obtained from the 
N3LO EM 500 and AV$_{18}$ interactions for the bare and the two different SRG resolution scales in 
Fig.~\ref{fig-model-dep} with and without the self-energy corrections. We only document the results for 
$\kf < 2.0 \,\fmi$, that is within the conservative estimate for the validity of the free space N3LO 
interactions. With a free spectrum (solid lines with filled symbols), there is not much difference in the 
relative error between the bare and the SRG evolved interactions except for $\kf$ close to $2.0 \,\fmi$. 
When the first order self energy correction is included, the error between the N3LO and AV$_{18}$ is quite large 
when the bare interactions are compared, but that decreases once the interactions are SRG evolved to
lower $\lambda$ values. All the same, compared to Fig.~\ref{fig-self-ener-SRG-errors}, the relative errors 
between the models are quite large and this can be attributed to the fact that the interactions are not phase shift 
equivalent (see Fig.~\ref{fig-srg-ps}). In fact, by comparing the N3LO interaction with a 
phenomenological one, we see the effects of free space phase shift inequivalence on the 
triplet gaps, which ties well with our motivation to study the gaps as a function of the SRG resolution 
scale instead.

\section{Conclusions}
\label{sect:summary}

The goal of this work is to understand the pairing physics of the triplet channel in pure neutron matter. 
We use SRG-evolved interactions as inputs as they preserve the bare phase shifts by construction and 
hence the phase shift inequivalence is factored out of the results and one can attribute the resulting 
dependence on the SRG resolution scale $\lambda$ to the missing many-body/medium effects. 

The zero temperature gaps are obtained by solving the angle-averaged BCS gap equation using the
numerically stable procedure of Khodel et al as well as by the stability method that locates the pole of the 
in-medium $T$ matrix. We also obtain the transition temperature using the Thouless criterion 
generalized for non-local interaction and verify that $\D\frac{\Delta(\kf)}{T_c} \sim 1.76$ at the BCS level. 

The gaps should be independent of the SRG resolution scale $\lambda$ and therefore, any $\lambda$ 
dependence is used as a tool to estimate the missing many-body/medium corrections. To this effect, we 
included the three-body effects at 
leading order via a density dependent effective two-body interaction. Including the three-body term at 
leading order does not change the two-body $\lambda$ dependence and further, we note via the 
dependence on the $3N$ cut-off that the short-range three-body effects are important. As a result, 
given these uncertainties with the leading order $3N$, we have not included the $3N$ corrections to the 
self-energy studies. 

The self-energy at first-order is included and this results in significant changes in the overall values of the 
gaps, gap closure as well as reduced $\lambda$ dependence. In fact, with the first order self-energy
correction we see trends similar to that observed in the literature when short-range correlations are 
included~\cite{Dong:2013sqa,Ding:2015tda,Ding:2016oxp}. However, we would like to emphasize that 
higher-order corrections to the self-energy are important. 

These results are encouraging at this point and it would be interesting to further investigate systematically, higher-order many-body/medium effects using the EFT approach and this is currently in progress. 

\begin{acknowledgements}
We would like thank Dick Furnstahl, Michael Urban, Kai Hebeler and Christian Drischler for useful 
comments and suggestions. We would like to acknowledge the contribution of Kai Hebeler who provided 
the three-body matrix elements and Christian Drischler who provided us the data for benchmarking our 
codes. We would like to thank Paolo Finelli, Stefano Maurizio and Wilhem Dickhoff who discussed with us 
their results and numerical implementation in detail. SR acknowledges support from the New Faculty 
Seed Grant from IIT Madras which provided the infrastructure to carry out the numerical calculations in 
this paper.
\end{acknowledgements}

\appendix*
\section{SRG Evolution}
\label{sect:appen1}

Similarity Renormalization Group~\cite{Bogner:2009bt,Bogner:2006pc} achieves the decoupling of low- 
and high-momentum modes using unitary transformations of the Hamiltonian. This has the unique 
feature of driving the high-momentum states towards the diagonal, which makes it different from the RG 
methods so far seen. The low-momentum effective potential obtained using Similarity Renormalization 
Group (SRG) denoted as $\vsrg$, preserves all the phase shifts with respect to the bare potential. 
	
Consider the following transformation on the hamiltonian
\beq
	H_s = U(s) H U^{\dagger}(s) \equiv T_{{\rm rel}} + V_s
	\label{srg.eqn1}
\eeq
where $U(s)$ and $U^{\dagger}(s)$ are unitary operators. Now 
\beq
	\D\frac{dH_s}{ds} = \frac{dU_s}{ds} H U^{\dagger}(s) + U(s) H \frac{dU^{\dagger}}{ds}.
	\label{srg.eqn2}
\eeq	
Defining $\eta(s) = \D\frac{dU(s)}{ds} U^{\dagger}(s)$, Eq.~(\ref{srg.eqn2}) becomes,
\beq
	\D\frac{dH_s}{ds} = \eta(s) H_s + H_s \eta^{\dagger}(s).
	\label{srg.eqn3}
\eeq
Using $U^{\dagger}(s) U(s) = 1$, one can establish that $\eta^{\dagger}(s) + \eta(s) = 0$, so that,   
Eq.~(\ref{srg.eqn3}) can be written as,
\beq
	\D\frac{dH_s}{ds} = \eta(s) \,H_s - H_s\, \eta(s) = \left[\eta(s), H_s \right].
	\label{srg.eqn6}
\eeq
We have the freedom to choose $\eta(s)$ and the canonical choice is~\cite{Bogner:2006pc}:
\beq
	\eta(s) = \left[T_{{\rm rel}}, H_s \right] = \D\frac{dU(s)}{ds}\, U^{\dagger}(s). 
	\label{srg.eqn7}
\eeq
The flow equation given by Eq~(\ref{srg.eqn6}) now reads:
\beq
	\D\frac{dH_s}{ds} = \left[\left[T_{{\rm rel}}, H_s \right], H_s \right].
	\label{srg.eqn8}
\eeq					
After some algebra one can show that Eq.~(\ref{srg.eqn8}) reduces to
\begin{multline}
	\D\frac{dV_s(k^\prime, k)}{ds} = -(k^2 - (k^\prime)^2)^2 \,V_s(k^\prime, k)\\ 
	+ \D\frac{2}{\pi} \int_0^{\lambda} q^2 dq \, (k^2 + (k^\prime)^2 - 2q^2)\, V_s(k^\prime, q)\, V_s(q, k)
	\label{srg.eqn9}
\end{multline}
Far away from the diagonal, the first term in Eq.~(\ref{srg.eqn9}) dominates, that is:
\beq
	\D\frac{dV_s(k^\prime, k)}{ds} \approx -(k^2 - (k^\prime)^2)^2 \,V_s(k^\prime, k),
\eeq
so that,
\beq
	V_s(k^\prime, k) \approx V_0(k^\prime, k) e^{-(k^2 - k^{\prime\,2})^2 s}.
	\label{srg.eqn10}
\eeq
Therefore it is clear that the off-diagonal elements are exponentially suppressed. Using the fact that 
$s^{-1/4}$ has the dimensions of momentum, one can define $\lambda = s^{-1/4}$, where $\lambda$ 
measures the spread of the off-diagonal strength.


\end{document}